\documentclass[fleqn,usenatbib]{mnras}

\usepackage{newtxtext,newtxmath}

\usepackage[T1]{fontenc}

\DeclareRobustCommand{\VAN}[3]{#2}
\let\VANthebibliography\thebibliography
\def\thebibliography{\DeclareRobustCommand{\VAN}[3]{##3}\VANthebibliography}

\usepackage{graphicx}	
\usepackage{amsmath}	
\usepackage{comment}	
\usepackage{soul}
\usepackage{xcolor}

\newcommand{\Ms}{M$_\odot$}

\newcommand{\bh}{\textit{Gaia} BH1}

\title[Dynamical 
formation of \textit{Gaia} BH1]
{Dynamical 
formation of \textit{Gaia} BH1 in a young star cluster}

\author[S. Rastello et al.]{
\newauthor Sara Rastello$^{1,2}$\thanks{E-mail: sara.rastello@unipd.it}, 
Giuliano Iorio$^{1,2,3}$\thanks{E-mail: giuliano.iorio@unipd.it}, 
Michela Mapelli$^{1,2,3}$\thanks{E-mail: michela.mapelli@unipd.it},
Manuel Arca-Sedda$^{1,4}$, Ugo N. Di Carlo $^{5}$, \newauthor Gast\'on J. Escobar$^{1,2}$,
Stefano Torniamenti$^{1,2,3}$, Tomer Shenar$^{6}$\\
$^{1}$Physics and Astronomy Department Galileo Galilei, University of Padova, Vicolo dell'Osservatorio 3, I--35122, Padova, Italy\\
$^{2}$INFN - Padova, Via Marzolo 8, I--35131 Padova, Italy\\
$^{3}$INAF - Osservatorio Astronomico di Padova, Vicolo dell'Osservatorio 5, I-35122 Padova, Italy\\
$^{4}$ Gran Sasso Science Institute (GSSI), 67100, L’Aquila, Italy\\
$^{5}$ Scuola Internazionale Superiore di Studi Avanzati (SISSA), Via Bonomea 265, I-34136 Trieste, Italy \\
$^{6}$ Anton Pannekoek Institute for Astronomy, University of Amsterdam, Postbus 94249, 1090 GE Amsterdam, The Netherlands\\
}

\date{Accepted XXX. Received YYY; in original form ZZZ}

\pubyear{2023}

\begin{document}
\label{firstpage}
\pagerange{\pageref{firstpage}--\pageref{lastpage}}
\maketitle

\begin{abstract}
\bh, the first quiescent black hole (BH) detected from {\it Gaia} data, poses a challenge to most binary evolution models: its current mass ratio is  $\approx{0.1}$, and its orbital period seems to be too long for a post-common envelope system and too short for a non-interacting binary system. Here, we explore the hypothesis that {\it Gaia} BH1 formed through dynamical interactions in a young star cluster (YSC). We study the properties of BH-main sequence (MS) binaries formed in YSCs with initial mass $3\times{}10^2-3\times{}10^4$~M$_\odot$ at solar metallicity, by means of $3.5\times{}10^4$ direct $N$-body simulations coupled with binary population synthesis. For comparison, we also run a sample of isolated binary stars with the same binary population synthesis code used in the dynamical models. We find that  BH-MS systems that form via dynamical exchanges populate the region corresponding to the main orbital properties of \bh{} (period, eccentricity, and masses). In contrast, none of our isolated binary systems matches the orbital period and MS mass of \bh{}. Our best matching \bh--like system forms via repeated dynamical exchanges and collisions involving the BH progenitor star, before it undergoes core collapse. YSCs are at least two orders of magnitude more efficient in forming \bh--like systems than isolated binary evolution.

\end{abstract}



\section{Introduction}
\label{sec:intro}

The zoo of black hole (BH) binaries has never been so rich. For several decades, a dozen X-ray binary systems with dynamical  measurement of the BH mass \citep{oezel2010,farr2011,miller2015} represented the main, if not the only, observational benchmark for theoretical models of BH formation.  
In 2015, the LIGO--Virgo collaboration started a revolution by detecting the merger of two BHs for the very first time \citep{abbottGW150914}. Since then, the number of gravitational-wave event candidates 
has grown rapidly, and is now approaching the 100 mark \citep{abbottGWTC2.1,abbottGWTC3}. In the last few years, a third family of observable BHs in binaries found its renaissance: quiescent BHs, i.e. X-ray silent BHs that are members of a binary system with a companion star.

Both astrometry and radial velocity measurements can help us to spot a quiescent BH \citep{trimble1969}. In the framework of a MUSE spectroscopic survey of 25 globular clusters (GCs), \cite{giesers2018} found evidence of a solar-mass  turn-off main-sequence (MS) star orbiting around an unseen companion of minimum mass $\sim{4}$ M$_\odot${}. Again with radial velocities, \cite{thompson2019} probed a binary system in the Galactic field, composed of a $\sim{3}$ M$_\odot$ giant star and a $\approx{3}$~M$_\odot$ dark companion. 
Furthermore, \cite{mahy2022} reported a $\approx 7$ \Ms{} BH
orbiting the Galactic O star HD 130298 ($\approx 24$ \Ms). 
\cite{shenar2022a} found spectroscopic evidence for the first extra-Galactic quiescent BH: VTFS243 is a young binary system in the Large Magellanic Cloud, comprising an O-type star of 25 M$_\odot$ and an unseen companion of at least 9 M$_\odot$.
A number of additional quiescent BH candidates are still debated, because they are consistent with alternative explanations \citep[e.g.,][]{casares2014,ribo2017,lennon2022,saracino2022,saracino2023}.

The \textit{Gaia} satellite yields the largest potential for constraining quiescent BHs, theoretical models suggesting that \textit{Gaia} will reveal several hundreds to thousands of quiescent BHs, mostly via astrometric measurements \citep{breivik2017,mashian2017,yalinewich2018,yamaguchi2018,shahaf2019,andrews2019,shikauchi2020,chawla2022,janssens2022,andrews2023,chakrabarti2022}. Recently, the \textit{Gaia} collaboration has published $\sim{3}\times{10^5}$ astrometric and spectroscopic binary stars in \textit{Gaia} Data Release 3 \citep{gaia2022,holl2022,halbwachs2022}. These data contain a number of quiescent BH candidates \citep{shion2022,shahaf2023,andrews2023b}. In particular, the two strongest candidates are \bh{} \citep{elbadry2023a} and BH2 \citep{elbadry2023b,tanikawa2023bh}.

\bh{} is a BH-MS binary system with estimated BH mass of about 10 M$_\odot$, and a solar-mass main-sequence companion \citep{elbadry2023a}. It has a current secondary-to-primary\textbf{\footnote{We define the primary star as the most massive object whilst the secondary is the less massive one.}} mass ratio $q\sim{0.1}$. If it formed in isolation, i.e. from a field binary system, the initial mass ratio should have been even more extreme $q\sim{0.01-0.05}$, below the minimum value currently assumed by most models ($q_\mathrm{min}=0.1$). 
This lower limit is potentially an artifact stemming from the difficulty of probing mass ratios more extreme than 0.1. However, things become more problematic when we also consider the current orbital period, $P_\mathrm{orb}=185.6$ days \citep{elbadry2023a}.

Roche-lobe filling binaries with a large donor-to-accretor mass ratio undergo dynamically unstable mass transfer (i.e., a common-envelope episode), according to mass-transfer stability criteria commonly adopted in population-synthesis codes \citep{hurley02}. In the standard $\alpha-$formalism \citep{webbink84}, the semi-major axis of the binary systems shrinks by orders of magnitude during a common envelope episode. The current semi-major axis of \bh{} is sufficiently short that the progenitor binary system must have gone through Roche lobe overflow. Given the large donor-to-accretor mass ratio, this mass transfer episode must have been unstable. If this was the case, the current orbital period should, however, be much shorter than its observed value \citep{breivik2017,elbadry2023a}.

On the one hand, the stability of mass transfer and dynamics of common envelope 
are still intensely debated problems  \citep{Ivanova2013,marchant2021,gallego-garcia2021,roepke2022}: it might be that \bh{} is a further alarm bell of the many limitations of mass-transfer models. On the other, \bh{} might have been the outcome of a merger in a triple system \citep{elbadry2023a}, or the result of dynamical assembly in a dense star cluster  \citep[e.g.,][]{shikauchi2020,tanikawa2023}. 
Binary-single star encounters in star clusters enable the formation of binary systems with exotic orbital properties \citep{heggie75,Portegies-Zwart10}. In particular, BHs are very effective in acquiring companions via dynamical exchanges, because they are more massive than most other stellar objects in clusters \citep{hills1980,mapelli16,dicarlo2019,Rastello2018,Trani2022}.

Here, we explore the formation of \bh--like systems in young star clusters (YSCs) with an age $\leq10^2$ Myr and initial stellar mass $3\times{}10^2-3\times{}10^4$ M$_\odot$, by means of direct $N$-body simulations. Such YSCs have very short two-body relaxation timescales ($\sim{10-100}$ Myr), which enhance gravitational encounters in the first few Myr of the star cluster life \citep{rastello2021}, and are fast disrupted in the Galactic tidal field \citep{torniamenti2022}.

Section~\ref{sec:methods} presents our simulations; Section~\ref{sec:results} summarizes our main results. In Section~\ref{sec:discussion}, we discuss the evolution of our best \bh-like candidate, the formation efficiency of such systems, and we compare our results with the isolated formation channel and  previous work. Finally, we lay down our Conclusions in Section~\ref{sec:conclusions}.

\begin{table}
	\centering
	\caption{Properties of \bh. \textit{From top to bottom}: 
     mass of the BH,  
	$m_{\rm{BH}}$ (\Ms); mass of the MS star, $m_{\rm{MS}}$ (\Ms); orbital 
	period, $P$ (days); orbital eccentricity, $e$; 
 Galactic location; environment metallicity $Z$, and reference paper.}
	\label{tab:bh1}
	\begin{tabular}{lc}
    \hline
        \multicolumn{2}{c}{\textbf{\bh}   \bf{     parameters}}                           \\ \hline
		$m_{\rm{BH}}$  (\Ms) & $9.6 \pm{} 0.18$    \\ 
		$m_{\rm{MS}}$ (\Ms)  & $0.93 \pm{} 0.05$   \\
		$P$ (days)                         & $185.59 \pm{} 0.05$ \\
		$e$                                 & $0.45 \pm{} 0.005$  \\
		Location                              & Milky Way disc    \\
		$Z$                                  & Near solar ([Fe/H] $= - 0.2$ )        \\
		Ref.                                & \cite{elbadry2023a}         \\ 
		\hline
	\end{tabular}
\end{table}


\section{Methods}
\label{sec:methods}

\subsection{Direct $N$-body simulations}
Here, we make use of the same direct $N$-body simulations that we presented in \citet{Rastello20,rastello2021} and \citet{dicarlo20}. We refer to these papers for further details. 
We performed our simulations with the code \textsc{nbody6++gpu} \citep{wang15}, the GPU parallel version of \textsc{nbody6}
\citep{Aarseth03}. \textsc{nbody6++gpu} makes use of a 4th-order Hermite integrator, individual
block time–steps \citep{makino92} and Kustaanheimo-Stiefel 
regularization for close encounters and few-body systems
\citep{stiefel65}.
 
We used our custom version of \textsc{nbody6++gpu}, which is interfaced with the binary population synthesis code \textsc{mobse} 
\citep{Mapelli17,giacobbo18,giacobbomapelli18}. \textsc{mobse} is a custom and upgraded version of \textsc{bse} \citep{hurley00,hurley02}, including up-to-date
prescriptions for stellar winds, pair instability, pulsational pair-instability, electron capture, and 
 core-collapse supernovae. 
In this set of simulations, we adopt the rapid core-collapse supernova model \citep{fryer12}, which enforces a mass gap between $2-5$~\Ms.
We 
use the same formalism as \cite{hurley02} for binary evolution (tides, mass transfer, common envelope and
GW orbital decay).  
In particular, we describe the common-envelope formalism with the $\alpha{}$ formalism, in which the free parameter 
$\alpha{}$  
represents the fraction of orbital energy that is available to unbind the common envelope.  
Higher values of $\alpha$ imply higher chances to successfully unbind the envelope and wider post-common envelope binaries.
Values of $\alpha{}>1$ are often used to take into account 
additional sources of energy not included in the $\alpha{}$ formalism \citep[e.g.,][]{klencki2021}.
In our simulations we use $\alpha{}=5$, consistent with \cite{giacobbomapelli18} and \cite{fragos2019}. In particular, \cite{fragos2019} retrieve a value of $\alpha=5$ from a set of hydrodynamical simulations of the common-envelope phase, while \cite{giacobbomapelli18} show that such high value of the common-envelope parameter is favoured by the merger rates inferred from the LIGO--Virgo collaboration (see also \citealt{sgalletta2023}).   
We randomly generate natal kicks of neutron stars  
from a Maxwellian velocity distribution with 
a one-dimensional root mean square velocity $\sigma{}=15$~km~s$^{-1}$ \citep{giacobbomapelli18}. 
For BHs, we adopt the same Maxwellian distribution and the kick magnitude is then lowered  by a factor depending on fallback efficiency, 
as described in \cite{fryer12}. This distribution yields lower kicks for BHs than estimated from BH X-ray binary systems, but still within their large uncertainties \citep[e.g.,][]{repetto2012,repetto2017,atri2019}.

We consider two sets of star cluster simulations with solar metallicity ($Z=0.02$). The first set was described in \cite{Rastello20} and consists of $3.3\times{}10^4$ clusters with mass  $300 <m_{\rm SC}/{\rm M}_\odot{}< 1000$ (see Table 1 in \citealt{Rastello20}). Hereafter, we will refer to them as low-mass (LM) clusters. 

The second-set was presented in \cite{dicarlo20} and consists of $2\times{}10^3$ star clusters with mass $10^{3} <m_{\rm SC}/{\rm M}_\odot{}< 3\times 10^{4}$ (see Table 1 in \citealt{dicarlo20}). Hereafter, we will refer to this second set as high-mass (HM) clusters. The HM clusters consist of two sub-samples: A) dense YSCs, with initial half-mass density $\rho_{\rm h}\ge{}3.4\times10^4$ M$_\odot$ pc$^{-3}$, and B) loose YSCs, 
 with initial half-mass density $1.5\times10^2 \le{} \rho_{\rm h}/$ (M$_\odot$ pc$^{-3}) \le{}3.4\times10^4$ and initial half-mass radius $r_{\rm h}=1.5$ pc. 

 In both LM and HM sets, the initial star cluster masses are sampled according to a power-law distribution $dN/dm_{\rm SC}\propto m_{\rm SC}^{-2}$ \citep{lada2003}. We used the software 
\textsc{McLuster} \citep{Kupper11} to generate the YSC initial conditions. 
We draw the initial positions and velocities of single stars and binary systems according to the fractal initial conditions of \textsc{McLuster}, starting from a homogeneous sphere and assuming a fractal parameter $D=1.6$ and a virial ratio $Q\equiv{K/|W|}=0.5$, where $K$ and $W$ are the total kinetic and potential energy of the cluster, respectively.  Fractal initial conditions mimic the clumpiness and asymmetry of YSCs and star forming regions \citep{ballone2020,ballone2021,torniamenti2021}. 

The initial mass function of stars in each cluster follows a \cite{kroupa2001} mass function in the range $0.1-150$ \Ms. 
Each YSC hosts initially $40$\% original binaries, whose mass ratios ($q$), periods ($P$) and eccentricities ($e$) are generated following the distributions reported by \cite{sana12}. In particular, orbital periods ($P$) follow the distribution
$\mathcal{P}(\Pi)\propto{}\Pi^{-0.55}$, where $\Pi\equiv{}\log_{10}(P/\mathrm{days})$
and $0.15\leq{}\Pi\leq{}6.7$, while binary eccentricities $e$
are randomly drawn from a distribution $\mathcal{P}(e)\propto{}e^{-0.42}$ with
$0\leq{}e<1$. Stars are paired up according to their mass following a distribution:
$\mathcal{P}(q)\propto{}q^{-0.1}$, where $q=m_2/m_1$ and $q \in [0.1,1.0]$ is the ratio of the mass between
the secondary and the primary star according to \cite{sana12}.
We assume that all stars with mass $\,m\,\ge{}5$~\Ms{} are members of binary systems, while stars with mass
$m\,<\,5$~\Ms{} are randomly paired until we reach a total binary fraction equal to $40$\%.

Our YSCs move on a circular orbit around the centre of the Milky Way at a distance of $8\,\mathrm{kpc}$ \citep{wang15} in a solar neighbourhood-like static external potential. We evolve each YSC for $100$ Myr.

\subsection{Binary population synthesis simulations}\label{sec:comparison}

For comparison with our dynamical binaries, we have run a sample of isolated binary systems with \textsc{mobse} \citep{Mapelli17,giacobbo18,giacobbomapelli18}, the very same  population-synthesis code that we used in our  dynamical simulations. Furthermore, we use the same setup of \textsc{mobse} for both dynamical and isolated simulations (e.g., the same common-envelope efficiency, mass transfer formalism, stellar wind model, and  natal kick model).
 
Finally, we use the original  binary stars present in the initial conditions of our star cluster simulations as initial conditions for the sample of isolated binary stars. These assumptions guarantee that we can do a fair comparison between isolated binaries and  dynamical simulations.

\section{Results} \label{sec:results} \label{res_dyn}
\begin{figure}
    \centering      
\includegraphics[width=0.5\textwidth]{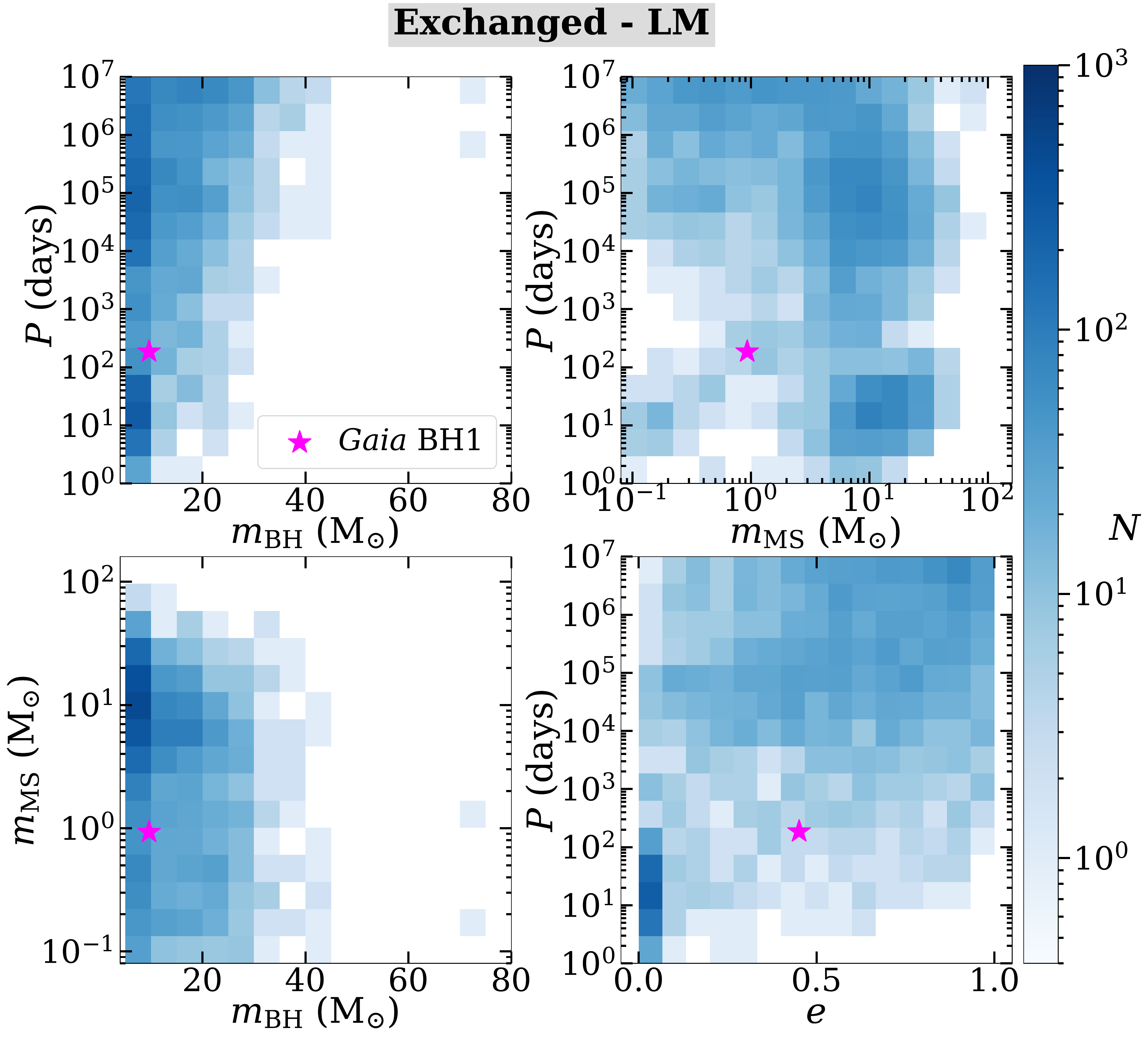}
    \caption{Orbital properties of  
    exchanged BH-MS binaries ejected from LM clusters, shown at the time of ejection. 
    \textit{Upper left}: BH mass $m_{\rm{BH}}$ (\Ms) versus binary period $P$ (days); \textit{upper right}: main-sequence (MS) star mass $m_{\rm{MS}}$ (\Ms) versus orbital period $P$ (days); \textit{lower left}: BH mass $m_{\rm{BH}}$ (\Ms) versus MS star mass $m_{\rm{MS}}$ (\Ms); \textit{lower right panel}: orbital eccentricity ($e$) versus orbital period $P$ (days). The colour map indicates the number ($N$) of systems per each bin. The magenta star refers to  
    \bh{} 
    \protect\citep{elbadry2023a}. One-$\sigma{}$ uncertainties (Table~\ref{tab:bh1}) are smaller than marker's size.}
    \label{fig:low-ej-ex}
\end{figure}

\begin{figure}
    \centering
    \includegraphics[width=0.5\textwidth]{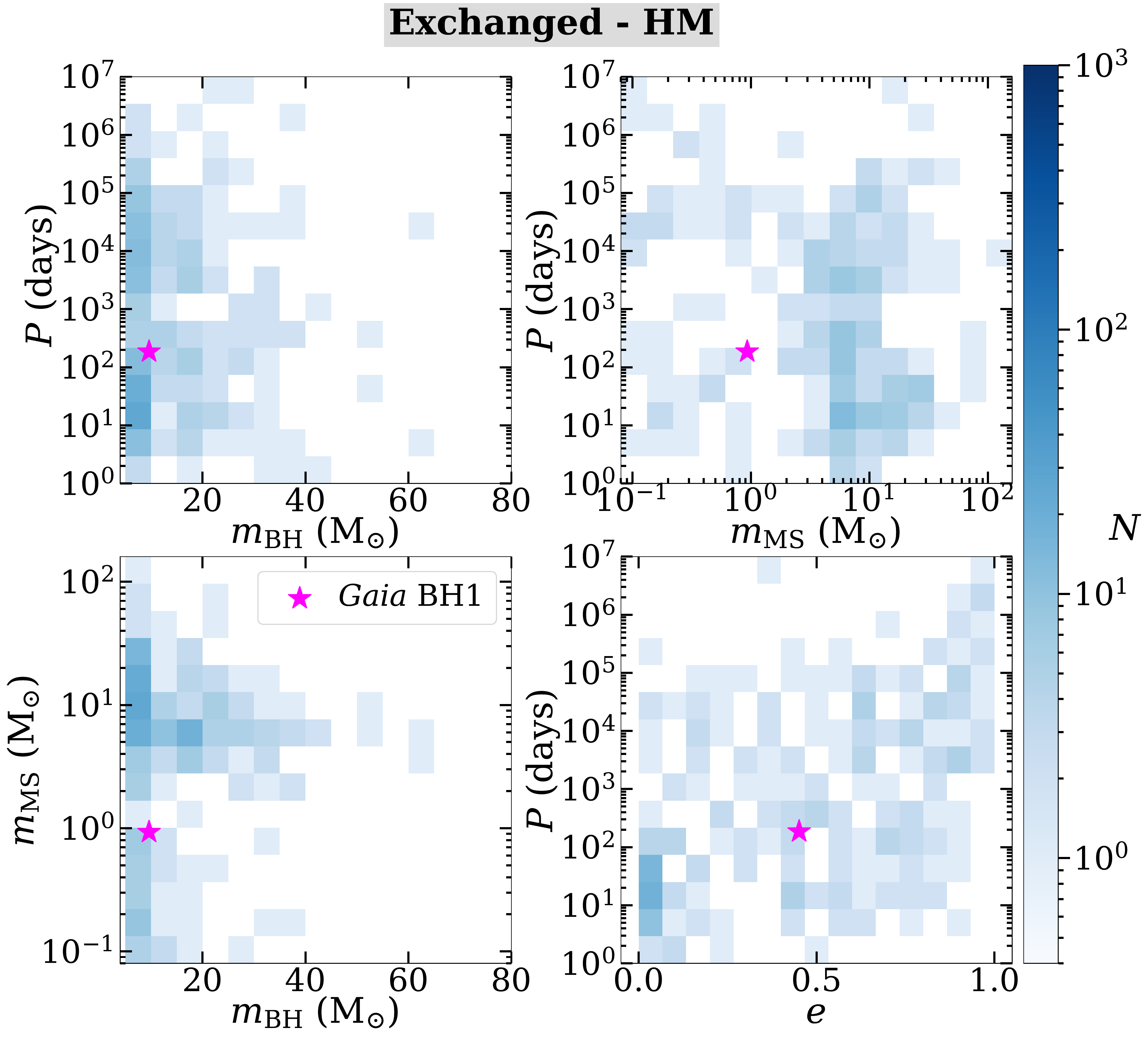}
    \caption{Same as Fig.~\ref{fig:low-ej-ex} but for
    HM clusters.} 
    \label{fig:hm_ex} 
\end{figure}

\begin{figure}    
\centering  
    \includegraphics[width=0.45\textwidth]{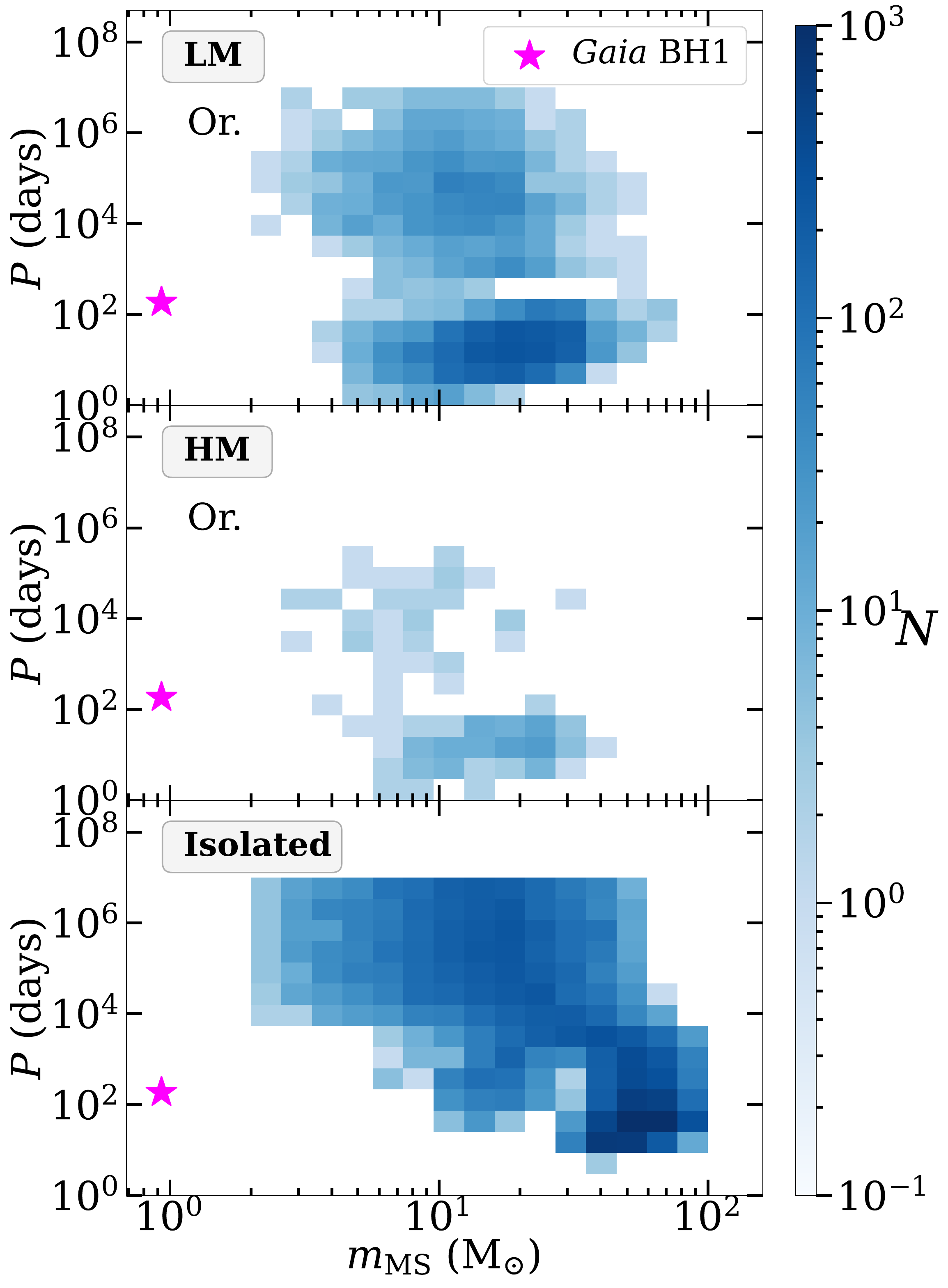} 
    \caption{Orbital period $P$ 
    versus MS mass $m_{\rm{MS}}$.
    The \textit{upper} and \textit{middle panels} refer to escaped original BH-MS binaries in LM and HM clusters, respectively. The \textit{lower panel} refers to isolated binaries (field binaries) at the time of their formation as BH-MS binaries. The magenta star refers to 
    \bh{} \protect\citep{elbadry2023a}. The 
    colour map indicates the number 
    of BH-MS binaries.}
    \label{fig:orig_iso}
\end{figure}

\bh{} is not a member of any star cluster nowadays. Hence, in our analysis, we only consider BH-MS binaries that are no longer members of their parent star cluster at the end of the simulation. Some of them have been ejected via dynamical recoil, others quietly evaporated from the star cluster. 
For simplicity, we will refer to all of them as ejected stars. In practice, we consider ejected all those stars and binary systems that  are at  a distance  larger than twice the cluster's tidal radius  
from the density centre of the YSC \citep{aarsethnb7,wang15}.

We simulated our star clusters for 100 Myr. \bh{} is likely an old system (several Gyr, \citealt{elbadry2023a}). Hence, it might have been ejected from its parent star cluster at a time  longer than 100 Myr. Simulating all of our YSCs for several Gyr is computationally prohibitive. Moreover, our tidal field model does not include tidal disc shocks and perturbers (e.g., giant molecular clouds), which cause premature disruption of star clusters \citep[e.g.,][]{gieles2006,kruijssen2015}. Hence, in Appendix~\ref{sec:app}, we consider not only the ejected BH-MS sytems but also the other systems, since most of them might be a field population at the age of \bh{}.

 Binary systems in clusters belong to two subgroups: \textit{original binaries} formed at cluster's birth, whereas \textit{exchanged binaries} assemble through dynamical encounters during cluster evolution.

Most of the BH-MS binaries produced by LM clusters are exchanged binaries ($\approx60$\%), whilst the remaining are original. 
About $\approx40$\% of all the BH-MS binaries formed in LM leave the cluster in the first 100 Myr, with $\approx45$\% ($\approx55$\%) of them being exchanged (original) binaries. As discussed in \citet[][section 3.1 and figure 1]{rastello2021}, this owes 
to the rapid expansion of the half-mass radius in LM clusters, which in turn causes a rapid drop of the central density and a continuous loss of stars over the simulated time.

Similarly, HM clusters form $\approx70$\% ($\approx 30$\%) exchanged (original) BH-MS binaries. However, HM clusters retain a larger fraction of binaries with respect to LM clusters. Only $7$\% of all formed BH-MS binaries escape from HM clusters during the simulation: 57\% of them are exchanged and 43\% original binaries. 

This happens because stars and binaries need a higher kinetic energy to exceed the escape velocity of HM clusters ($V_{\rm{esc}}\sim 10$ km/s), and HM clusters  survive longer against tidal disruption than LM clusters \citep[see e.g.,][]{torniamenti2022}. 

Figure~\ref{fig:low-ej-ex} 
shows the main physical properties of  exchanged BH-MSs binaries formed in LM clusters, at the time of their ejection.  
We can distinguish three subgroups of binaries, based on the orbital period and mass of the MS star:
\begin{itemize}
    \item[i)] short/intermediate period--high MS mass binaries: $1<P/\rm{days}<10^{2}$ and $2<m_{\rm{MS}}/\rm{M}_{\odot}<50$. Binaries in this group are mostly circular (80\% with $e\approx0$) and most of them (90\%) host a BH with mass lower than 10 \Ms; 
    \item[ii)] short/intermediate period-low MS mass binaries ($1 \lessapprox P/\rm{days} \lessapprox 10^{3}$ and $0.1 \lessapprox m_{\rm{MS}}/\rm{M}_{\odot} \lessapprox 2$). Binaries belonging to this group have period and MS mass that correlate almost linearly, indeed they appear as a strip in the period-MS mass plane bound by the lines $P\approx20 \left(\frac{m_{\rm{MS}}}{\rm{M}_{\odot}}\right) \ \rm{days}$ and $P\approx 400 \left(\frac{m_{\rm{MS}}}{\rm{M}_{\odot}}\right) \ \rm{days}$. The eccentricity of such binaries lies in the range $0<e<0.6$. They mainly host (62\%) BHs with mass $<10$~\Ms, even if the BH mass distribution extends up to 30 \Ms;
    \item[iii)] long-period binaries ($10^{4}<P/\mathrm{days}< 10^{7}$ and $0.1 < m_{\rm{MS}}/\rm{M}_{\odot}<50$). These BH-MS binaries have preferentially eccentric orbits ($>52$\% have $e>0.7$) and host massive BHs ($\sim50$\% binaries in which $m_{\rm{BH}}>10$ \Ms). 
\end{itemize}
Groups (i) and (ii) form preferentially through the pairing of a MS star and a BH's progenitor that underwent at least one common-envelope episode. This explains why most binaries in groups (i) and (ii) exhibit nearly circular orbits. The ejection of these binaries is driven by the natal kick as the primary component turns into a BH. Roughly 20\% of exchanged BH-MS binaries belong to group (i), while only 3--4\% binaries fall in group (ii). Binaries in group (iii) with a low-mass MS star ($m_{\rm{MS}} \approx 1-2$~\Ms) form through the pair up of a BH and a MS star, while those with a heavier MS component form through the pair up of the BH's progenitor star and the MS star. Most ejected exchanged binaries (75\%) belong to group (iii).

Figure~\ref{fig:hm_ex} shows the properties of ejected BH-MS binaries in HM clusters. 

This population is much sparser in HM clusters compared to LM clusters because of the lower number. However, the distribution of period--MS mass in HM clusters is comparable to that of LM clusters, except for long-period binaries that are easily disrupted in more massive clusters.

Finally, the lower panel of Fig.~\ref{fig:orig_iso} shows the orbital period--MS mass of isolated BH-MS systems, compared to original binaries in LM (upper) and HM clusters (middle).

Short-period isolated binaries ($1< \mathrm{P}/$days$<10^{2}$) form through common-envelope evolution, while long-period binaries ($10^{2}< P/$ days $<10^{6}$) either do not interact or interact only through stable mass transfer.

Although original binaries in clusters share the same initial conditions as 
binaries evolved in isolation (Section~\ref{sec:comparison}), the final distribution of BH-MS properties shows some differences. In fact, dynamical interactions
favour the additional shrinking of tight binaries producing a population of  short period BH-MS binaries ($P<1$ days), 
which does not exist in isolated BH-MS binaries (lower panel of Fig. \ref{fig:orig_iso}). 
In contrast, dynamical interactions tend to disrupt wide binaries reducing the number of long-period BH-MS binaries. This effect is particularly relevant in the HM clusters, where  
dynamical interactions produce a strong suppression of 
long-period BH-MS systems.

The minimum mass-ratio ($q_{\rm{min}}=0.1$) assumed  
in the initial conditions of original binaries  limits the minimum mass of the MS component to $\approx 2 \ \mathrm{M}_\odot$ in both the LM and HM clusters.

\subsection{\bh{} parameter space} \label{sec:intbh}

The magenta stars in Figures \ref{fig:low-ej-ex}, \ref{fig:hm_ex}, and \ref{fig:orig_iso} indicate the observed properties of \bh{} \citep{elbadry2023a}.
The apparent lack of original and isolated binaries with properties similar to \bh{} (Figure \ref{fig:orig_iso}) is mostly due to the choice of the initial minimum mass ratio ($q_\mathrm{min}=0.1$).  
As a consequence, the lowest mass ($m_{\rm{MS}}\approx 2.2 $ \Ms) of the MS star in a BH-MS binary is too high to match \bh. 

Here, we have assumed $q_{\rm min}=0.1$ even in the isolated binary simulations in order to ensure a fair comparison with our YSC simulations. We will relax this assumption about $q_{\rm min}$ in future works.
However, the production of \bh{}-like systems in original and isolated binaries remains challenging, even if we  relax the limit on the initial mass ratio  (Section~\ref{sec:iso}).

In contrast, dynamical interactions in LM and HM clusters 
produce exchanged BH-MS systems with properties similar to \bh. Such systems belong to the sub-population of exchanged BH-MS binaries that we labelled as group (ii) in Section~\ref{res_dyn}, and form from the pair-up of a low-mass MS star and a massive BH progenitor (mass ratio $q\lesssim0.1$). 

As previously mentioned, only about $3-4\%$ of the ejected exchanged binaries form through this channel. This fraction 
becomes even smaller $(<1\%$) if we consider only  binary systems with BH and MS star mass close to the values of \bh{}  (Table \ref{tab:bh1} and Section~\ref{sec:eff}).

\section{Discussion}
\label{sec:discussion}

\subsection{Dynamical origin of a \textit{Gaia} BH1-like binary system} \label{sec:bh1dyn}
 
\bh{}  could have been formed dynamically in dense star clusters \citep{elbadry2023a}, 
such as  
globular clusters \citep{kremer2018},  
YSCs and open clusters \citep{shikauchi2020}. 
Given the high metallicity of the MS companion and the location of \bh{} in 
the Galactic thin disc, the formation scenario in a globular cluster is disfavoured. Still, \bh{} could have formed in a YSC from a long-period massive binary system \citep{fujii11} or from a dynamically assembled binary system that shrinks via close encounters \citep{Banerjee18a,Banerjee18b} and is either ejected 
or  
evaporates after  
cluster dissolution \citep{Schoettler2019}.

\begin{figure}    
\centering  
    \includegraphics[width=0.40\textwidth]{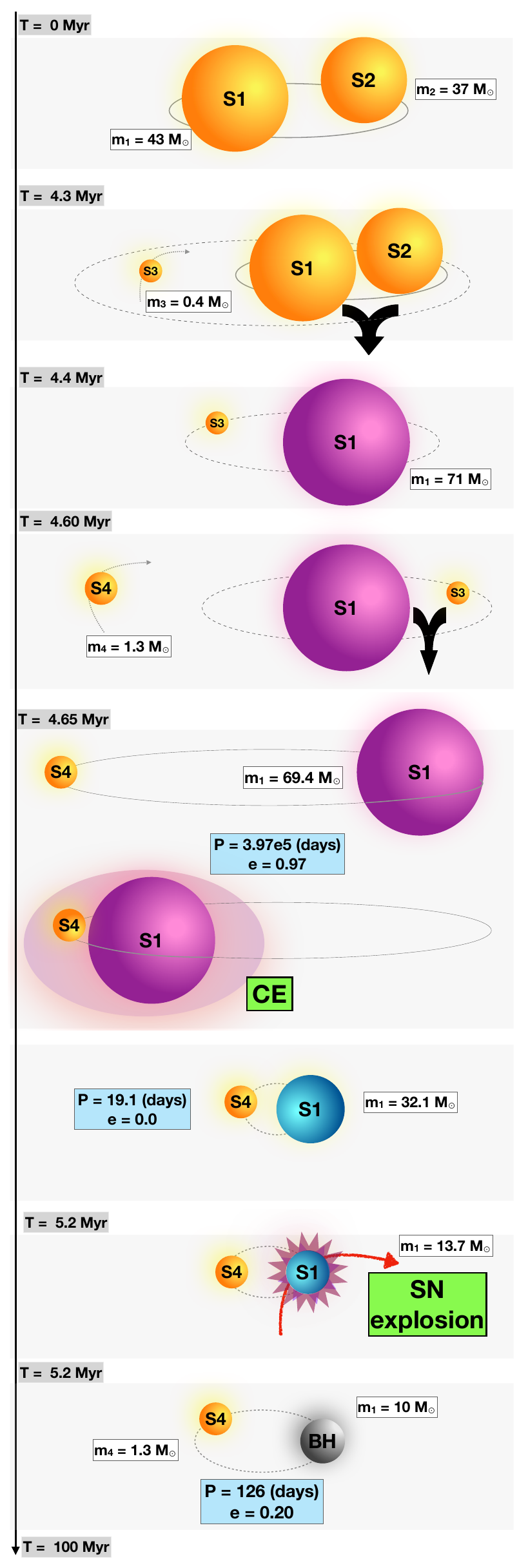}   
    \caption{History of the simulated BH-MS system that best matches the properties of \bh{}.
    The simulated system forms in a LM cluster ($m_{\rm SC}\approx{320}$ M$_\odot$). The timeline in Myr is indicated on the left-hand side of the cartoon. The colours refer to different stellar types: orange for MS stars, magenta for core-helium burning giant stars, cyan for pure-He stars, and grey for BHs.  
    When S1 and S4 form a binary system, the orbital period ($P$) and eccentricity ($e$) of the binary are shown
    in the blue box.}  
\label{fig:cartoon}
\end{figure}

Figure~\ref{fig:cartoon} sketches the formation history of our  
simulated 
BH-MS system 
that best matches \bh{}. 
The simulated BH-MS system forms in a relatively LM cluster with $m_{\rm{SC}}\approx 320$ \Ms.
Its progenitors are an original binary composed of two quite massive stars (S1 and S2, with mass $m_{1} \approx 43$ \Ms{} and $m_{2} \approx 37$ \Ms{}, respectively) and a single low-mass MS star (S4).

At $\approx{4.3}$ Myr, the original S1-S2 binary system is perturbed by a low mass MS star (S3) which triggers the binary collision. The merger product is a MS star that  
then evolves into a core helium burning star (S1) of $\approx 71$ \Ms{}, bound to S3. 

At 4.6 Myr, the system composed of the massive star S1 and the low-mass MS star S3 is furthermore perturbed by another MS star (S4) with $m_{4} \approx 1.3$ \Ms{}. 
The dynamical perturbation 
triggers a merger between S1 and S3; the collision product remains bound to the perturber S4 in a eccentric orbit ($e\approx0.97$). 
At $t\approx5$ Myr, the binary separation at periastron ($\approx 2000 $ R$_{\odot}$) is smaller than the sum of the stellar radii ($\approx 3000 $ R$_{\odot}$). Thus, the system undergoes a common-envelope evolution that removes the envelope of S1 producing a pure-He star.

After the common-envelope episode, the binary orbit is almost circular with a short period ($P \approx 20 $ days). Then, the star S1 experiences a significant mass loss  which reduce its mass to $m_{1} \approx 13.7$ \Ms. Finally, at $t\approx5.2$ Myr, S1 undergoes a core-collapse supernova, leaving a BH of $\approx 10$ \Ms{} and receiving a low natal kick, which does not destroy the binary system.
The binary composed of the BH (S1) and the MS star (S4) is our best \bh{}-like system, which after $\approx{9}$ Myr escapes from the parent cluster with $e=0.2$ and an orbital period $P\sim 130$ days.

The evolution shown in Fig. \ref{fig:cartoon} is 
a common formation channel of BH-MS binaries with physical parameters in the range of \bh{} in our simulated LM and HM clusters. In our clusters, dynamical encounters and exchanges occur mostly between stars and compact-object progenitors \citep{dicarlo20,Rastello20,rastello2021,torniamenti2022}, because of their short two-body relaxation time (a few Myr). 
Our LM clusters are nearly disrupted by the tidal field at the end of our simulations (100 Myr, \citealt{torniamenti2022}). Hence, we can exclude that additional \bh--like systems could form at later stages of LM cluster evolution.
In contrast, most of the HM clusters are still marginally  affected by the tidal field at the end of the simulation \citep{dicarlo20}: we cannot exclude that additional \bh{}-like systems could have formed through 
dynamical exchanges with a BH intruder in our HM clusters, if we 
had continued the simulation after 100 Myr.

\subsection{Comparison with the isolated formation channel} \label{sec:iso}

The lower panel of Fig.~\ref{fig:orig_iso}  shows that no \bh--like systems form in our isolated binaries. 

This primarily stems from the assumption of $q_\mathrm{min}=0.1$ in the initial conditions. In contrast, dynamical interactions facilitate the formation of exchanged binaries across the entire range of mass ratios. In fact, the \bh--like system we described in Sec. \ref{sec:bh1dyn} originates from an exchanged BH-MS binary with $q\approx 0.02$ (Fig. \ref{fig:cartoon}).

The assumption of $q_\mathrm{min}=0.1$ for original and isolated binaries comes from the range assumed by  \cite{sana12} in their fit to $\mathcal{P}(q)$. 
If this is 
a strong constraint  for the formation of   binary systems, the result presented  in Figs. \ref{fig:low-ej-ex} and \ref{fig:hm_ex} 
supports the conclusion that \bh{} can form from neither isolated nor original binaries. 
However, 
the lack of observed binaries with $q<0.1$ could be attributed to observational biases \citep[e.g.,][]{moe2015,moe17,tokovinin2020}.  
Hence, the mass ratio alone is not enough to
distinguish between a dynamical or an isolated formation of \bh. 

Previous studies have investigated the formation of BH-MS systems through isolated binary evolution 
without imposing restrictions on the range of the initial mass ratios \citep[see e.g.][]{breivik2017,Shao2019,Wiktorowicz2020,chawla2022,shikauchi2022,shikauchi23}.   Despite large variations in BH-MS properties resulting from different simulation assumptions (such as metallicity, natal kick model, and wind prescriptions), these studies consistently indicate that it is not possible to form \bh--like systems without adopting very high common-envelope efficiencies ($\alpha>5$, see, e.g., \citealt{elbadry2023a}). 
 
For 
$\alpha<5$, the common-envelope phase leads to the production of binaries that are too tight ($P\lesssim 1$ days). 
In contrast, when higher common-envelope efficiencies are considered, the formation pathways for \bh--like systems (e.g., type 2 channel in \citealt{shikauchi23}) resemble those observed in our clusters (Sections~\ref{sec:intbh} and \ref{sec:bh1dyn}).

Dynamical interactions in  clusters generate a larger variety of stellar and binary properties at the onset of common envelope. 
Specifically, 
dynamical interactions 
can trigger star--star collisions,  
leading to the formation of stars with unusual properties  
(in terms of, e.g., envelope-core mass ratio, or evolutionary stage at the onset of common envelope, see e.g. 
\citealt{ivanova2002,schneider2019,schneider2020,costa2022,ballone2023}).
Furthermore,  
star clusters 
enhance the formation of highly eccentric binaries ($e>0.9$). 
For example, Fig. \ref{fig:cartoon} shows that the BH progenitor of our best \bh{}--like system 
experiences a sequence of stellar collisions before pairing up with the low-mass companion and triggering a common-envelope phase with an highly eccentric orbit. The differences in the properties at the onset of common envelope between isolated and exchanged binaries 
explain why our simulations can produce \bh--like systems without assuming very high $\alpha$ ($>10$).

There is an alternative scenario that would allow that formation of \bh{} from an isolated binary system: if the progenitor star of the BH never evolves into a red super-giant star (for example because of efficient mass loss), its radius remains $<1000$~R$_\odot$ even in the late evolutionary stages. In such case, the progenitor star remains sufficiently compact that \bh{} might originate from a non-interacting binary system, without any common-envelope phase. This scenario is not allowed in our simulations, because the stellar tracks assumed in \textsc{mobse} and other \textsc{bse}-like codes \citep{hurley02} commonly evolve through the red-giant phase, even at high metallicity. However, in other stellar evolution models, the BH progenitor remains quite compact at high $Z$, avoiding the red super-giant phase (see, e.g., figure 7 of \citealt{iorio2023} and figure 2 of \citealt{agrawal2022}). Furthermore, most stellar evolution models are known to over-predict the population of red giant stars with respect to observations \citep[e.g.,][]{Gilkis2021,Romagnolo2022}. Thus, we will explore alternative stellar evolution models for \bh{} in future works.

\subsection{Formation efficiency} \label{sec:eff}

\begin{table}
\centering
\begin{tabular}{@{}c|cc|cc@{}}
\hline
\textbf{}   & \multicolumn{2}{c}{\textbf{all BH-MSs}}  & \multicolumn{2}{c}{\textbf{ejected BH-MSs}}     \\ 
\textbf{}   & \textbf{$\eta_{\rm{tot}}$}  & \textbf{$\eta_{\rm BH1}$} & \textbf{$\eta_{\rm{tot}}$}  & \textbf{$\eta_{\rm BH1}$} \\ 
 \textbf{}   &  ($10^{-4}$ M$_{\odot}^{-1}$)  &    ($10^{-7}$ M$_{\odot}^{-1}$)  &  ($10^{-4}$ M$_{\odot}^{-1}$)   &  ($10^{-7}$ M$_{\odot}^{-1}$)   \\ \hline
\textbf{LM} & $10.73^{+0.13}_{-0.13}$      & $2.09^{+2.32}_{-1.31}$  & $4.33^{+0.08}_{-0.08}$  &  $0.95^{+1.74}_{-0.75}$    \\
& & & &  \\
\textbf{HM} & $8.13^{+0.17}_{-0.16}$       & $2.08^{+3.80}_{-1.64}$    &$0.58^{+0.05}_{-0.04}$  &  $2.08^{+3.80}_{-1.64}$    \\ \hline
\end{tabular}
\caption{Formation efficiency ($\eta$) of BH-MS systems (first 2 columns) and ejected BH-MS binaries (last 2 columns) in LM and HM clusters. From left to right the columns in each sample refer to i) $\eta_{\rm{tot}}$: formation efficiency of  BH-MS systems, 
ii) $\eta_{\rm{BH1}}$: formation efficiency of BH-MS systems 
that fall into the \bh{}-selection window, 
as described in Section~\ref{sec:eff}. The efficiency $\eta$ is assumed to be the rate of a Poisson process and we constrain its posterior distribution given the number of objects, $P(\eta|N)$, using the Bayes theorem. We report the median of the posterior distribution as the fiducial value, while the  reported upper and lower bounds define the 90\% of the posterior distribution. We calculate the efficiency considering the cumulative number of formed BH-MS systems (each binary is counted just once at the moment of its creation).}
\label{tab:efficiency}
\end{table}

To quantify the ability of a YSC to form a population of BH-MS binaries, we define a formation efficiency ($\eta$) as:
\begin{equation}
\centering
\eta = N_{\rm{BH-MS}}\, /\, m_{\rm{SC}}^{\rm{tot}},
\end{equation}
where $N_{\rm{BH-MS}}$ is the number of BH-MS binaries formed and $m_{\rm{SC}}^{\rm{tot}}$ is the total simulated initial stellar mass in star clusters: $m_{\rm{LM}}^{\rm{tot}} \approx 10^7$~M$_{\odot}$ for  LM clusters and $m_{\rm{HM}}^{\rm{tot}} \approx 8\times 10^6$~M$_{\odot}$ for HM clusters.
 
Table~\ref{tab:efficiency} shows the formation efficiency, distinguishing between all BH-MS binaries (left-hand side columns in Table \ref{tab:efficiency}) and the subclass of binaries ejected from the parent cluster (right-hand side columns of Table \ref{tab:efficiency}).  
In calculating $\eta{}$, we 
group together 
original and exchanged BH-MS binaries. 
Furthermore, we estimate the formation efficiency limiting the range of physical parameters to \bh{}-like systems considering the following selection cuts:
$ 0.5< m_{\rm{MS}}/\mathrm{M}_\odot  <1.5 $, $8< m_{\rm{BH}}/\mathrm{M}_\odot  <12$, $80 < P/\mathrm{days} < 500 $, and $0.2 < e < 0.7$.

The formation efficiency of BH-MS systems is  comparable in LM and HM clusters. 
If we restrict our analysis to ejected binaries, LM clusters are about $\approx7$ times more efficient in forming  
ejected binaries than HM clusters. 
This stems from the different efficiency of HM and LM clusters in retaining stars and binaries.  
Within the uncertainties, there are no significant differences between LM and HM clusters in forming ejected BH-MS binaries with mass comparable to \bh{} ($\eta_{\rm{BH1}}$).

The total formation efficiency of isolated BH-MS binaries is 
$15.9\pm0.2 \times 10^{-4} \ \mathrm{M}^{-1}_\odot$, 
comparable to that found for LM and HM clusters. 
In contrast, the formation efficiency \bh{}-like systems 
is zero for isolated binaries (Fig.~\ref{fig:orig_iso}). 
We refer to \cite{DiCarlo2023} for a comprehensive model of the formation and detection rate of \bh{} systems in the Milky Way.

\subsection{Comparison with previous work}

The ejected exchanged binaries presented in this work have similar properties 
to those found by \citet{shikauchi2020} in their open cluster simulations. In particular, we populate mostly the period--MS mass region of their group I and group III. Their group I systems form through three-body encounters and binary-single interactions between stars which then experience a common-envelope episode before BH formation, while their group~III systems consist of binaries formed via direct capture in a three-body and/or binary-single encounter.

Recently, \cite{tanikawa2023} conducted a study focused on the dynamical formation channel of \bh-like systems, carrying out $N$-body simulations of YSCs with $m_{\rm{SC}}\sim10^3$~M$_\odot$ at $Z = 0.02,$ 0.01, and 0.005, up to 500 Myr with 100\% primordial binaries. 
Selecting binaries with parameters in the range $m_{\rm{MS}} <1.1$~M$_\odot$, $100 < P/\mathrm{days}< 2000$ and $0.3 < e < 0.9 $, they found that YSCs are about three orders of magnitude more efficient in forming \bh-like systems than isolated binaries ($\eta\sim10^{-5} \ \mathrm{M}^{-1}_\odot$ and $\eta\sim10^{-8} \ \mathrm{M}^{-1}_\odot$, respectively). Their best \bh{}-like system is a dynamical binary, found in a cluster at $Z=0.005$, built after a series of exchanges and dynamical captures that lead to the final formation 
of a triple system which is then ejected from the parent cluster.

Using the same 
definition of \bh{}-like systems as in \cite{tanikawa2023}, the \bh{}-like formation efficiency in our simulations (both LM and HM clusters) is $\eta\sim10^{-6} \ \mathrm{M}^{-1}_\odot$, about one order of magnitude lower than the one found by  \cite{tanikawa2023}. 
This difference can be attributed to the initial conditions of our simulations compared  to \citet{tanikawa2023}, in terms of fraction of original binaries (40\% versus 100\% in our simulations and in \citealt{tanikawa2023}, respectively), limit on the mass ratio ($q_{\rm{min}}=0.1$ versus no lower limit), cluster's initial fractality, common-envelope efficiency ($\alpha=5$ versus $\alpha=3$), natal kick and stellar metallicity ($Z=0.02$ versus $Z=0.02$, 0.001, and 0.005).

\section{Summary and conclusions}
\label{sec:conclusions}
We studied the population of BH-MS binaries formed in low-mass (LM, $3\times10^2-10^3$ M$_\odot$) and high-mass (HM, $10^3-3\times{}10^4$ M$_\odot$) young star clusters at Solar metallicity, by means of a set of $3.5\times{}10^4$ direct $N$-body simulations with population-synthesis calculations. Our aim is to investigate the formation of \bh{}-like systems \citep{elbadry2023a}. 

\bh{} is in the Galactic field, but its  current low mass ratio ($q\approx{0.1}$) and  orbital period ($P=186$ days) are 
difficult to explain with isolated binary evolution. In particular, its current orbital period requires that its progenitor binary system underwent at least one episode of Roche-lobe overflow, unless we assume that the BH progenitor was a rather compact star ($<1000$ R$_\odot$) for its entire life. According to current models of mass-transfer stability, the extreme mass ratio of \bh{} implies that the mass-transfer episode was unstable, triggering a common-envelope phase. However, this scenario would have led to a much shorter orbital period than 186 days, unless we assume an unrealistically high value of the common-envelope efficiency parameter (\citealt{elbadry2023a} report $\alpha=14$ as their best value).

Dynamical interactions in a star cluster add several degrees of freedom to this picture. The BH or its progenitor star might have paired up dynamically with the low-mass MS star at some point during the evolution of the parent cluster \citep{shikauchi2020,tanikawa2023}. In such case, the semi-major axis and eccentricity of the binary system are set by the dynamical exchange, by subsequent dynamical interactions and by the natal kick, rather than by the initial properties of the progenitor binary star. 
The BH-MS binary system might then have become a field system after being dynamically ejected, or because the cluster dissolved into the tidal field of the Milky Way.

We find that 
exchanged BH-MS systems (i.e., 
BH-MS systems that form via dynamical exchanges) in our $N$-body simulations populate the region corresponding to the main orbital properties of \bh{}: the orbital period $P$, eccentricity $e$, mass of the MS companion $m_{\rm MS}$, and mass of the BH $m_{\rm BH}$ (Figs~\ref{fig:low-ej-ex} and \ref{fig:hm_ex}). In contrast, 
original BH-MS systems 
(i.e., those binary systems that we injected in the initial conditions of our star cluster simulations) cannot explain the orbital period of \bh{}. 
 We also ran a comparison sample of isolated binary systems with our population-synthesis code \textsc{mobse}. None of our isolated binary systems matches the orbital period and MS mass of \bh{} (Fig.~\ref{fig:orig_iso}).

The formation efficiency of ejected \bh{}-like systems in our $N$-body simulations is relatively high: $\sim{2\times{}10^{-7}}$~M$_\odot^{-1}$ 
in both LM and HM clusters.

From our dynamical simulations, we selected the system that best matches the properties of \bh{} and studied its evolution (Fig.~\ref{fig:cartoon}). This system forms in one of our LM clusters with mass $m_{\rm SC}=320$~M$_\odot$. 
The BH progenitor is a massive star ($\sim 70$ \Ms), which forms from the collision of two  
MS stars (43 and 37 M$_\odot$, respectively), members of an original binary system. This collision is triggered by a binary-single encounter with a low-mass MS star (0.4 M$_\odot$), which then becomes bound to the collision product. At 4.6 Myr, the BH progenitor and the 0.4 M$_\odot$ star collide and merge, because of another binary-single encounter with a 1.3 M$_\odot$ MS star, which then remains bound to the collision product. The new binary system undergoes a common-envelope phase, after which the primary member of the binary system undergoes a core-collapse supernova. The supernova explosion leaves a BH with mass 10 M$_\odot$, still bound to the 1.3 M$_\odot$ MS companion. The natal kick 
induces the ejection of the binary system from the cluster and leads to the final orbital properties: $P=126$ days and $e=0.2$. 
Hence, our best \bh{}-like candidate forms through a combination of dynamical encounters, 
stellar evolution, and core-collapse supernova physics: dynamical encounters lead to the pair up of the low-mass MS and the BH progenitor star, stellar evolution triggers the common-envelope episode, while the natal kick shapes the final orbital properties of the BH-MS system.


\section*{Acknowledgements} 
GE, GI, MM, SR, ST and UNDC acknowledge financial support from the European Research Council for the ERC Consolidator grant DEMOBLACK, under contract no. 770017.
MAS acknowledges funding from the European Union’s Horizon 2020 research and innovation programme under the Marie Skłodowska-Curie grant agreement No. 101025436 (project GRACE-BH). TS was supported by the European Union's Horizon 2020 Marie Skłodowska-Curie grant No.\ 101024605.
This work benefited from support of 
the Munich Institute for Astro-, Particle and BioPhysics 
(MIAPbP), funded by the Deutsche Forschungsgemeinschaft (DFG, German Research Foundation) 
under Germany's Excellence Strategy – EXC-2094 – 390783311. 
We acknowledge the CINECA-INFN agreement, for the availability of high performance computing resources and support.

\section*{Data availability}
The data underlying this article will be shared on reasonable request to the corresponding authors.



\bibliographystyle{mnras}
\bibliography{main_mnras} 

\begin{thebibliography}{}
\makeatletter
\relax
\def\mn@urlcharsother{\let\do\@makeother \do\$\do\&\do\#\do\^\do\_\do\%\do\~}
\def\mn@doi{\begingroup\mn@urlcharsother \@ifnextchar [ {\mn@doi@}
  {\mn@doi@[]}}
\def\mn@doi@[#1]#2{\def\@tempa{#1}\ifx\@tempa\@empty \href
  {http://dx.doi.org/#2} {doi:#2}\else \href {http://dx.doi.org/#2} {#1}\fi
  \endgroup}
\def\mn@eprint#1#2{\mn@eprint@#1:#2::\@nil}
\def\mn@eprint@arXiv#1{\href {http://arxiv.org/abs/#1} {{\tt arXiv:#1}}}
\def\mn@eprint@dblp#1{\href {http://dblp.uni-trier.de/rec/bibtex/#1.xml}
  {dblp:#1}}
\def\mn@eprint@#1:#2:#3:#4\@nil{\def\@tempa {#1}\def\@tempb {#2}\def\@tempc
  {#3}\ifx \@tempc \@empty \let \@tempc \@tempb \let \@tempb \@tempa \fi \ifx
  \@tempb \@empty \def\@tempb {arXiv}\fi \@ifundefined
  {mn@eprint@\@tempb}{\@tempb:\@tempc}{\expandafter \expandafter \csname
  mn@eprint@\@tempb\endcsname \expandafter{\@tempc}}}

\bibitem[\protect\citeauthoryear{Aarseth}{Aarseth}{2003}]{Aarseth03}
Aarseth S.~J.,  2003, Gravitational N-Body Simulations: Tools and Algorithms.
Cambridge Monographs on Mathematical Physics, Cambridge University Press

\bibitem[\protect\citeauthoryear{{Aarseth}}{{Aarseth}}{2012}]{aarsethnb7}
{Aarseth} S.~J.,  2012, \mn@doi [\mnras] {10.1111/j.1365-2966.2012.20666.x},
  \href {http://adsabs.harvard.edu/abs/2012MNRAS.422..841A} {422, 841}

\bibitem[\protect\citeauthoryear{{Abbott} et~al.,}{{Abbott}
  et~al.}{2016}]{abbottGW150914}
{Abbott} B.~P.,  et~al., 2016, \mn@doi [\prl] {10.1103/PhysRevLett.116.061102},
  \href {https://ui.adsabs.harvard.edu/abs/2016PhRvL.116f1102A} {116, 061102}

\bibitem[\protect\citeauthoryear{{Abbott} et~al.,}{{Abbott}
  et~al.}{2021a}]{abbottGWTC2.1}
{Abbott} R.,  et~al., 2021a, \mn@doi [arXiv e-prints]
  {10.48550/arXiv.2108.01045}, \href
  {https://ui.adsabs.harvard.edu/abs/2021arXiv210801045T} {p. arXiv:2108.01045}

\bibitem[\protect\citeauthoryear{{Abbott} et~al.,}{{Abbott}
  et~al.}{2021b}]{abbottGWTC3}
{Abbott} R.,  et~al., 2021b, \mn@doi [arXiv e-prints]
  {10.48550/arXiv.2111.03606}, \href
  {https://ui.adsabs.harvard.edu/abs/2021arXiv211103606T} {p. arXiv:2111.03606}

\bibitem[\protect\citeauthoryear{{Agrawal}, {Sz{\'e}csi}, {Stevenson},
  {Eldridge}  \& {Hurley}}{{Agrawal} et~al.}{2022}]{agrawal2022}
{Agrawal} P.,  {Sz{\'e}csi} D.,  {Stevenson} S.,  {Eldridge} J.~J.,   {Hurley}
  J.,  2022, \mn@doi [\mnras] {10.1093/mnras/stac930}, \href
  {https://ui.adsabs.harvard.edu/abs/2022MNRAS.512.5717A} {512, 5717}

\bibitem[\protect\citeauthoryear{{Andrew}, {Penoyre}, {Belokurov}, {Evans}  \&
  {Oh}}{{Andrew} et~al.}{2022}]{shion2022}
{Andrew} S.,  {Penoyre} Z.,  {Belokurov} V.,  {Evans} N.~W.,   {Oh} S.,  2022,
  \mn@doi [\mnras] {10.1093/mnras/stac2532}, \href
  {https://ui.adsabs.harvard.edu/abs/2022MNRAS.516.3661A} {516, 3661}

\bibitem[\protect\citeauthoryear{{Andrews}, {Breivik}  \&
  {Chatterjee}}{{Andrews} et~al.}{2019}]{andrews2019}
{Andrews} J.~J.,  {Breivik} K.,   {Chatterjee} S.,  2019, \mn@doi [\apj]
  {10.3847/1538-4357/ab441f}, \href
  {https://ui.adsabs.harvard.edu/abs/2019ApJ...886...68A} {886, 68}

\bibitem[\protect\citeauthoryear{{Andrews}, {Taggart}  \& {Foley}}{{Andrews}
  et~al.}{2022}]{andrews2023b}
{Andrews} J.~J.,  {Taggart} K.,   {Foley} R.,  2022, \mn@doi [arXiv e-prints]
  {10.48550/arXiv.2207.00680}, \href
  {https://ui.adsabs.harvard.edu/abs/2022arXiv220700680A} {p. arXiv:2207.00680}

\bibitem[\protect\citeauthoryear{{Andrews}, {Breivik}, {Chawla}, {Rodriguez}
  \& {Chatterjee}}{{Andrews} et~al.}{2023}]{andrews2023}
{Andrews} J.~J.,  {Breivik} K.,  {Chawla} C.,  {Rodriguez} C.~L.,
  {Chatterjee} S.,  2023, \mn@doi [\apj] {10.3847/1538-4357/acbb5f}, \href
  {https://ui.adsabs.harvard.edu/abs/2023ApJ...946..111A} {946, 111}

\bibitem[\protect\citeauthoryear{{Atri} et~al.,}{{Atri}
  et~al.}{2019}]{atri2019}
{Atri} P.,  et~al., 2019, \mn@doi [\mnras] {10.1093/mnras/stz2335}, \href
  {https://ui.adsabs.harvard.edu/abs/2019MNRAS.489.3116A} {489, 3116}

\bibitem[\protect\citeauthoryear{{Ballone}, {Mapelli}, {Di Carlo},
  {Torniamenti}, {Spera}  \& {Rastello}}{{Ballone} et~al.}{2020}]{ballone2020}
{Ballone} A.,  {Mapelli} M.,  {Di Carlo} U.~N.,  {Torniamenti} S.,  {Spera} M.,
    {Rastello} S.,  2020, arXiv e-prints, \href
  {https://ui.adsabs.harvard.edu/abs/2020arXiv200110003B} {p. arXiv:2001.10003}

\bibitem[\protect\citeauthoryear{{Ballone}, {Torniamenti}, {Mapelli}, {Di
  Carlo}, {Spera}, {Rastello}, {Gaspari}  \& {Iorio}}{{Ballone}
  et~al.}{2021}]{ballone2021}
{Ballone} A.,  {Torniamenti} S.,  {Mapelli} M.,  {Di Carlo} U.~N.,  {Spera} M.,
   {Rastello} S.,  {Gaspari} N.,   {Iorio} G.,  2021, \mn@doi [\mnras]
  {10.1093/mnras/staa3763}, \href
  {https://ui.adsabs.harvard.edu/abs/2021MNRAS.501.2920B} {501, 2920}

\bibitem[\protect\citeauthoryear{{Ballone}, {Costa}, {Mapelli}, {MacLeod},
  {Torniamenti}  \& {Pacheco-Arias}}{{Ballone} et~al.}{2023}]{ballone2023}
{Ballone} A.,  {Costa} G.,  {Mapelli} M.,  {MacLeod} M.,  {Torniamenti} S.,
  {Pacheco-Arias} J.~M.,  2023, \mn@doi [\mnras] {10.1093/mnras/stac3752},
  \href {https://ui.adsabs.harvard.edu/abs/2023MNRAS.519.5191B} {519, 5191}

\bibitem[\protect\citeauthoryear{{Banerjee}}{{Banerjee}}{2018a}]{Banerjee18a}
{Banerjee} S.,  2018a, \mn@doi [\mnras] {10.1093/mnras/stx2347}, \href
  {https://ui.adsabs.harvard.edu/abs/2018MNRAS.473..909B} {473, 909}

\bibitem[\protect\citeauthoryear{{Banerjee}}{{Banerjee}}{2018b}]{Banerjee18b}
{Banerjee} S.,  2018b, \mn@doi [\mnras] {10.1093/mnras/sty2608}, \href
  {https://ui.adsabs.harvard.edu/abs/2018MNRAS.481.5123B} {481, 5123}

\bibitem[\protect\citeauthoryear{{Breivik}, {Chatterjee}  \&
  {Larson}}{{Breivik} et~al.}{2017}]{breivik2017}
{Breivik} K.,  {Chatterjee} S.,   {Larson} S.~L.,  2017, \mn@doi [\apjl]
  {10.3847/2041-8213/aa97d5}, \href
  {https://ui.adsabs.harvard.edu/abs/2017ApJ...850L..13B} {850, L13}

\bibitem[\protect\citeauthoryear{{Casares}, {Negueruela}, {Rib{\'o}}, {Ribas},
  {Paredes}, {Herrero}  \& {Sim{\'o}n-D{\'\i}az}}{{Casares}
  et~al.}{2014}]{casares2014}
{Casares} J.,  {Negueruela} I.,  {Rib{\'o}} M.,  {Ribas} I.,  {Paredes} J.~M.,
  {Herrero} A.,   {Sim{\'o}n-D{\'\i}az} S.,  2014, \mn@doi [\nat]
  {10.1038/nature12916}, \href
  {https://ui.adsabs.harvard.edu/abs/2014Natur.505..378C} {505, 378}

\bibitem[\protect\citeauthoryear{{Chakrabarti} et~al.,}{{Chakrabarti}
  et~al.}{2023}]{chakrabarti2022}
{Chakrabarti} S.,  et~al., 2023, \mn@doi [\aj] {10.3847/1538-3881/accf21},
  \href {https://ui.adsabs.harvard.edu/abs/2023AJ....166....6C} {166, 6}

\bibitem[\protect\citeauthoryear{{Chawla}, {Chatterjee}, {Breivik}, {Moorthy},
  {Andrews}  \& {Sanderson}}{{Chawla} et~al.}{2022}]{chawla2022}
{Chawla} C.,  {Chatterjee} S.,  {Breivik} K.,  {Moorthy} C.~K.,  {Andrews}
  J.~J.,   {Sanderson} R.~E.,  2022, \mn@doi [\apj] {10.3847/1538-4357/ac60a5},
  \href {https://ui.adsabs.harvard.edu/abs/2022ApJ...931..107C} {931, 107}

\bibitem[\protect\citeauthoryear{{Costa}, {Ballone}, {Mapelli}  \&
  {Bressan}}{{Costa} et~al.}{2022}]{costa2022}
{Costa} G.,  {Ballone} A.,  {Mapelli} M.,   {Bressan} A.,  2022, \mn@doi
  [\mnras] {10.1093/mnras/stac2222}, \href
  {https://ui.adsabs.harvard.edu/abs/2022MNRAS.516.1072C} {516, 1072}

\bibitem[\protect\citeauthoryear{{Di Carlo}, {Giacobbo}, {Mapelli}, {Pasquato},
  {Spera}, {Wang}  \& {Haardt}}{{Di Carlo} et~al.}{2019}]{dicarlo2019}
{Di Carlo} U.~N.,  {Giacobbo} N.,  {Mapelli} M.,  {Pasquato} M.,  {Spera} M.,
  {Wang} L.,   {Haardt} F.,  2019, \mn@doi [\mnras] {10.1093/mnras/stz1453},
  \href {https://ui.adsabs.harvard.edu/abs/2019MNRAS.487.2947D} {487, 2947}

\bibitem[\protect\citeauthoryear{{Di Carlo} et~al.,}{{Di Carlo}
  et~al.}{2020}]{dicarlo20}
{Di Carlo} U.~N.,  et~al., 2020, \mn@doi [\mnras] {10.1093/mnras/staa2286},
  \href {https://ui.adsabs.harvard.edu/abs/2020MNRAS.498..495D} {498, 495}

\bibitem[\protect\citeauthoryear{{Di Carlo}, {Rodriguez}  \& {Breivik}}{{Di
  Carlo} et~al.}{2023}]{DiCarlo2023}
{Di Carlo} Ugo Niccol{\`o}~Agrawal P.,  {Rodriguez} C.~L.,   {Breivik} K.,
  2023, \mn@doi [arXiv e-prints] {10.48550/arXiv.2306.13121}, \href
  {https://ui.adsabs.harvard.edu/abs/2023arXiv230613121D} {p. arXiv:2306.13121}

\bibitem[\protect\citeauthoryear{{El-Badry} et~al.,}{{El-Badry}
  et~al.}{2023a}]{elbadry2023a}
{El-Badry} K.,  et~al., 2023a, \mn@doi [\mnras] {10.1093/mnras/stac3140}, \href
  {https://ui.adsabs.harvard.edu/abs/2023MNRAS.518.1057E} {518, 1057}

\bibitem[\protect\citeauthoryear{{El-Badry} et~al.,}{{El-Badry}
  et~al.}{2023b}]{elbadry2023b}
{El-Badry} K.,  et~al., 2023b, \mn@doi [\mnras] {10.1093/mnras/stad799}, \href
  {https://ui.adsabs.harvard.edu/abs/2023MNRAS.521.4323E} {521, 4323}

\bibitem[\protect\citeauthoryear{{Farr}, {Sravan}, {Cantrell}, {Kreidberg},
  {Bailyn}, {Mandel}  \& {Kalogera}}{{Farr} et~al.}{2011}]{farr2011}
{Farr} W.~M.,  {Sravan} N.,  {Cantrell} A.,  {Kreidberg} L.,  {Bailyn} C.~D.,
  {Mandel} I.,   {Kalogera} V.,  2011, \mn@doi [\apj]
  {10.1088/0004-637X/741/2/103}, \href
  {https://ui.adsabs.harvard.edu/abs/2011ApJ...741..103F} {741, 103}

\bibitem[\protect\citeauthoryear{{Fragos}, {Andrews}, {Ramirez-Ruiz}, {Meynet},
  {Kalogera}, {Taam}  \& {Zezas}}{{Fragos} et~al.}{2019}]{fragos2019}
{Fragos} T.,  {Andrews} J.~J.,  {Ramirez-Ruiz} E.,  {Meynet} G.,  {Kalogera}
  V.,  {Taam} R.~E.,   {Zezas} A.,  2019, \mn@doi [\apjl]
  {10.3847/2041-8213/ab40d1}, \href
  {https://ui.adsabs.harvard.edu/abs/2019ApJ...883L..45F} {883, L45}

\bibitem[\protect\citeauthoryear{{Fryer}, {Belczynski}, {Wiktorowicz},
  {Dominik}, {Kalogera}  \& {Holz}}{{Fryer} et~al.}{2012}]{fryer12}
{Fryer} C.~L.,  {Belczynski} K.,  {Wiktorowicz} G.,  {Dominik} M.,  {Kalogera}
  V.,   {Holz} D.~E.,  2012, \mn@doi [\apj] {10.1088/0004-637X/749/1/91}, \href
  {https://ui.adsabs.harvard.edu/abs/2012ApJ...749...91F} {749, 91}

\bibitem[\protect\citeauthoryear{{Fujii} \& {Portegies Zwart}}{{Fujii} \&
  {Portegies Zwart}}{2011}]{fujii11}
{Fujii} M.~S.,  {Portegies Zwart} S.,  2011, \mn@doi [Science]
  {10.1126/science.1211927}, \href
  {https://ui.adsabs.harvard.edu/abs/2011Sci...334.1380F} {334, 1380}

\bibitem[\protect\citeauthoryear{{Gaia Collaboration} et~al.,}{{Gaia
  Collaboration} et~al.}{2022}]{gaia2022}
{Gaia Collaboration} et~al., 2022, \mn@doi [arXiv e-prints]
  {10.48550/arXiv.2206.05595}, \href
  {https://ui.adsabs.harvard.edu/abs/2022arXiv220605595G} {p. arXiv:2206.05595}

\bibitem[\protect\citeauthoryear{{Gallegos-Garcia}, {Berry}, {Marchant}  \&
  {Kalogera}}{{Gallegos-Garcia} et~al.}{2021}]{gallego-garcia2021}
{Gallegos-Garcia} M.,  {Berry} C. P.~L.,  {Marchant} P.,   {Kalogera} V.,
  2021, \mn@doi [\apj] {10.3847/1538-4357/ac2610}, \href
  {https://ui.adsabs.harvard.edu/abs/2021ApJ...922..110G} {922, 110}

\bibitem[\protect\citeauthoryear{{Giacobbo} \& {Mapelli}}{{Giacobbo} \&
  {Mapelli}}{2018}]{giacobbomapelli18}
{Giacobbo} N.,  {Mapelli} M.,  2018, \mn@doi [\mnras] {10.1093/mnras/sty1999},
  \href {https://ui.adsabs.harvard.edu/abs/2018MNRAS.480.2011G} {480, 2011}

\bibitem[\protect\citeauthoryear{{Giacobbo}, {Mapelli}  \& {Spera}}{{Giacobbo}
  et~al.}{2018}]{giacobbo18}
{Giacobbo} N.,  {Mapelli} M.,   {Spera} M.,  2018, \mn@doi [\mnras]
  {10.1093/mnras/stx2933}, \href
  {https://ui.adsabs.harvard.edu/abs/2018MNRAS.474.2959G} {474, 2959}

\bibitem[\protect\citeauthoryear{{Gieles}, {Portegies Zwart}, {Baumgardt},
  {Athanassoula}, {Lamers}, {Sipior}  \& {Leenaarts}}{{Gieles}
  et~al.}{2006}]{gieles2006}
{Gieles} M.,  {Portegies Zwart} S.~F.,  {Baumgardt} H.,  {Athanassoula} E.,
  {Lamers} H.~J.~G.~L.~M.,  {Sipior} M.,   {Leenaarts} J.,  2006, \mn@doi
  [\mnras] {10.1111/j.1365-2966.2006.10711.x}, \href
  {https://ui.adsabs.harvard.edu/abs/2006MNRAS.371..793G} {371, 793}

\bibitem[\protect\citeauthoryear{{Giesers} et~al.,}{{Giesers}
  et~al.}{2018}]{giesers2018}
{Giesers} B.,  et~al., 2018, \mn@doi [\mnras] {10.1093/mnrasl/slx203}, \href
  {https://ui.adsabs.harvard.edu/abs/2018MNRAS.475L..15G} {475, L15}

\bibitem[\protect\citeauthoryear{{Gilkis}, {Shenar}, {Ramachandran}, {Jermyn},
  {Mahy}, {Oskinova}, {Arcavi}  \& {Sana}}{{Gilkis} et~al.}{2021}]{Gilkis2021}
{Gilkis} A.,  {Shenar} T.,  {Ramachandran} V.,  {Jermyn} A.~S.,  {Mahy} L.,
  {Oskinova} L.~M.,  {Arcavi} I.,   {Sana} H.,  2021, \mn@doi [\mnras]
  {10.1093/mnras/stab383}, \href
  {https://ui.adsabs.harvard.edu/abs/2021MNRAS.503.1884G} {503, 1884}

\bibitem[\protect\citeauthoryear{{Halbwachs} et~al.,}{{Halbwachs}
  et~al.}{2022}]{halbwachs2022}
{Halbwachs} J.-L.,  et~al., 2022, \mn@doi [arXiv e-prints]
  {10.48550/arXiv.2206.05726}, \href
  {https://ui.adsabs.harvard.edu/abs/2022arXiv220605726H} {p. arXiv:2206.05726}

\bibitem[\protect\citeauthoryear{{Heggie}}{{Heggie}}{1975}]{heggie75}
{Heggie} D.~C.,  1975, \mn@doi [\mnras] {10.1093/mnras/173.3.729}, \href
  {https://ui.adsabs.harvard.edu/abs/1975MNRAS.173..729H} {173, 729}

\bibitem[\protect\citeauthoryear{{Hills} \& {Fullerton}}{{Hills} \&
  {Fullerton}}{1980}]{hills1980}
{Hills} J.~G.,  {Fullerton} L.~W.,  1980, \mn@doi [\aj] {10.1086/112798}, \href
  {https://ui.adsabs.harvard.edu/abs/1980AJ.....85.1281H} {85, 1281}

\bibitem[\protect\citeauthoryear{{Holl} et~al.,}{{Holl}
  et~al.}{2022}]{holl2022}
{Holl} B.,  et~al., 2022, \mn@doi [arXiv e-prints] {10.48550/arXiv.2206.05439},
  \href {https://ui.adsabs.harvard.edu/abs/2022arXiv220605439H} {p.
  arXiv:2206.05439}

\bibitem[\protect\citeauthoryear{{Hurley}, {Pols}  \& {Tout}}{{Hurley}
  et~al.}{2000}]{hurley00}
{Hurley} J.~R.,  {Pols} O.~R.,   {Tout} C.~A.,  2000, \mn@doi [\mnras]
  {10.1046/j.1365-8711.2000.03426.x}, \href
  {https://ui.adsabs.harvard.edu/abs/2000MNRAS.315..543H} {315, 543}

\bibitem[\protect\citeauthoryear{{Hurley}, {Tout}  \& {Pols}}{{Hurley}
  et~al.}{2002}]{hurley02}
{Hurley} J.~R.,  {Tout} C.~A.,   {Pols} O.~R.,  2002, \mn@doi [\mnras]
  {10.1046/j.1365-8711.2002.05038.x}, \href
  {https://ui.adsabs.harvard.edu/abs/2002MNRAS.329..897H} {329, 897}

\bibitem[\protect\citeauthoryear{{Iorio} et~al.,}{{Iorio}
  et~al.}{2023}]{iorio2023}
{Iorio} G.,  et~al., 2023, \mn@doi [\mnras] {10.1093/mnras/stad1630}, \href
  {https://ui.adsabs.harvard.edu/abs/2023MNRAS.tmp.1606I} {}

\bibitem[\protect\citeauthoryear{{Ivanova}, {Podsiadlowski}  \&
  {Spruit}}{{Ivanova} et~al.}{2002}]{ivanova2002}
{Ivanova} N.,  {Podsiadlowski} P.,   {Spruit} H.,  2002, \mn@doi [\mnras]
  {10.1046/j.1365-8711.2002.05543.x}, \href
  {https://ui.adsabs.harvard.edu/abs/2002MNRAS.334..819I} {334, 819}

\bibitem[\protect\citeauthoryear{{Ivanova} et~al.,}{{Ivanova}
  et~al.}{2013}]{Ivanova2013}
{Ivanova} N.,  et~al., 2013, \mn@doi [\aapr] {10.1007/s00159-013-0059-2}, \href
  {https://ui.adsabs.harvard.edu/abs/2013A&ARv..21...59I} {21, 59}

\bibitem[\protect\citeauthoryear{{Janssens} et~al.,}{{Janssens}
  et~al.}{2022}]{janssens2022}
{Janssens} S.,  et~al., 2022, \mn@doi [\aap] {10.1051/0004-6361/202141866},
  \href {https://ui.adsabs.harvard.edu/abs/2022A&A...658A.129J} {658, A129}

\bibitem[\protect\citeauthoryear{{Klencki}, {Nelemans}, {Istrate}  \&
  {Chruslinska}}{{Klencki} et~al.}{2021}]{klencki2021}
{Klencki} J.,  {Nelemans} G.,  {Istrate} A.~G.,   {Chruslinska} M.,  2021,
  \mn@doi [\aap] {10.1051/0004-6361/202038707}, \href
  {https://ui.adsabs.harvard.edu/abs/2021A&A...645A..54K} {645, A54}

\bibitem[\protect\citeauthoryear{{Kremer}, {Ye}, {Chatterjee}, {Rodriguez}  \&
  {Rasio}}{{Kremer} et~al.}{2018}]{kremer2018}
{Kremer} K.,  {Ye} C.~S.,  {Chatterjee} S.,  {Rodriguez} C.~L.,   {Rasio}
  F.~A.,  2018, \mn@doi [\apjl] {10.3847/2041-8213/aab26c}, \href
  {https://ui.adsabs.harvard.edu/abs/2018ApJ...855L..15K} {855, L15}

\bibitem[\protect\citeauthoryear{{Kroupa}}{{Kroupa}}{2001}]{kroupa2001}
{Kroupa} P.,  2001, \mn@doi [\mnras] {10.1046/j.1365-8711.2001.04022.x}, \href
  {http://adsabs.harvard.edu/abs/2001MNRAS.322..231K} {322, 231}

\bibitem[\protect\citeauthoryear{{Kruijssen}}{{Kruijssen}}{2015}]{kruijssen2015}
{Kruijssen} J.~M.~D.,  2015, \mn@doi [\mnras] {10.1093/mnras/stv2026}, \href
  {https://ui.adsabs.harvard.edu/abs/2015MNRAS.454.1658K} {454, 1658}

\bibitem[\protect\citeauthoryear{{K{\"u}pper}, {Maschberger}, {Kroupa}  \&
  {Baumgardt}}{{K{\"u}pper} et~al.}{2011}]{Kupper11}
{K{\"u}pper} A. H.~W.,  {Maschberger} T.,  {Kroupa} P.,   {Baumgardt} H.,
  2011, \mn@doi [\mnras] {10.1111/j.1365-2966.2011.19412.x}, \href
  {https://ui.adsabs.harvard.edu/abs/2011MNRAS.417.2300K} {417, 2300}

\bibitem[\protect\citeauthoryear{{Lada} \& {Lada}}{{Lada} \&
  {Lada}}{2003}]{lada2003}
{Lada} C.~J.,  {Lada} E.~A.,  2003, \mn@doi [\araa]
  {10.1146/annurev.astro.41.011802.094844}, \href
  {http://adsabs.harvard.edu/abs/2003ARA%26A..41...57L} {41, 57}

\bibitem[\protect\citeauthoryear{{Lennon}, {Dufton}, {Villase{\~n}or}, {Evans},
  {Langer}, {Saxton}, {Monageng}  \& {Toonen}}{{Lennon}
  et~al.}{2022}]{lennon2022}
{Lennon} D.~J.,  {Dufton} P.~L.,  {Villase{\~n}or} J.~I.,  {Evans} C.~J.,
  {Langer} N.,  {Saxton} R.,  {Monageng} I.~M.,   {Toonen} S.,  2022, \mn@doi
  [\aap] {10.1051/0004-6361/202142413}, \href
  {https://ui.adsabs.harvard.edu/abs/2022A&A...665A.180L} {665, A180}

\bibitem[\protect\citeauthoryear{{Mahy} et~al.,}{{Mahy}
  et~al.}{2022}]{mahy2022}
{Mahy} L.,  et~al., 2022, \mn@doi [\aap] {10.1051/0004-6361/202243147}, \href
  {https://ui.adsabs.harvard.edu/abs/2022A&A...664A.159M} {664, A159}

\bibitem[\protect\citeauthoryear{{Makino} \& {Aarseth}}{{Makino} \&
  {Aarseth}}{1992}]{makino92}
{Makino} J.,  {Aarseth} S.~J.,  1992, \pasj, \href
  {https://ui.adsabs.harvard.edu/abs/1992PASJ...44..141M} {44, 141}

\bibitem[\protect\citeauthoryear{{Mapelli}}{{Mapelli}}{2016}]{mapelli16}
{Mapelli} M.,  2016, \mn@doi [\mnras] {10.1093/mnras/stw869}, \href
  {https://ui.adsabs.harvard.edu/abs/2016MNRAS.459.3432M} {459, 3432}

\bibitem[\protect\citeauthoryear{{Mapelli}}{{Mapelli}}{2017}]{Mapelli17}
{Mapelli} M.,  2017, \mn@doi [\mnras] {10.1093/mnras/stx304}, \href
  {https://ui.adsabs.harvard.edu/abs/2017MNRAS.467.3255M} {467, 3255}

\bibitem[\protect\citeauthoryear{{Marchant}, {Pappas}, {Gallegos-Garcia},
  {Berry}, {Taam}, {Kalogera}  \& {Podsiadlowski}}{{Marchant}
  et~al.}{2021}]{marchant2021}
{Marchant} P.,  {Pappas} K. M.~W.,  {Gallegos-Garcia} M.,  {Berry} C. P.~L.,
  {Taam} R.~E.,  {Kalogera} V.,   {Podsiadlowski} P.,  2021, \mn@doi [\aap]
  {10.1051/0004-6361/202039992}, \href
  {https://ui.adsabs.harvard.edu/abs/2021A&A...650A.107M} {650, A107}

\bibitem[\protect\citeauthoryear{{Mashian} \& {Loeb}}{{Mashian} \&
  {Loeb}}{2017}]{mashian2017}
{Mashian} N.,  {Loeb} A.,  2017, \mn@doi [\mnras] {10.1093/mnras/stx1410},
  \href {https://ui.adsabs.harvard.edu/abs/2017MNRAS.470.2611M} {470, 2611}

\bibitem[\protect\citeauthoryear{{Miller} \& {Miller}}{{Miller} \&
  {Miller}}{2015}]{miller2015}
{Miller} M.~C.,  {Miller} J.~M.,  2015, \mn@doi [\physrep]
  {10.1016/j.physrep.2014.09.003}, \href
  {https://ui.adsabs.harvard.edu/abs/2015PhR...548....1M} {548, 1}

\bibitem[\protect\citeauthoryear{{Moe} \& {Di Stefano}}{{Moe} \& {Di
  Stefano}}{2015}]{moe2015}
{Moe} M.,  {Di Stefano} R.,  2015, \mn@doi [\apj]
  {10.1088/0004-637X/801/2/113}, \href
  {https://ui.adsabs.harvard.edu/abs/2015ApJ...801..113M} {801, 113}

\bibitem[\protect\citeauthoryear{{Moe} \& {Di Stefano}}{{Moe} \& {Di
  Stefano}}{2017}]{moe17}
{Moe} M.,  {Di Stefano} R.,  2017, \mn@doi [\apjs] {10.3847/1538-4365/aa6fb6},
  \href {https://ui.adsabs.harvard.edu/abs/2017ApJS..230...15M} {230, 15}

\bibitem[\protect\citeauthoryear{{O\"ezel}, {Psaltis}, {Narayan}  \&
  {McClintock}}{{O\"ezel} et~al.}{2010}]{oezel2010}
{O\"ezel} F.,  {Psaltis} D.,  {Narayan} R.,   {McClintock} J.~E.,  2010,
  \mn@doi [\apj] {10.1088/0004-637X/725/2/1918}, \href
  {https://ui.adsabs.harvard.edu/abs/2010ApJ...725.1918O} {725, 1918}

\bibitem[\protect\citeauthoryear{{Portegies Zwart}, {McMillan}  \&
  {Gieles}}{{Portegies Zwart} et~al.}{2010}]{Portegies-Zwart10}
{Portegies Zwart} S.~F.,  {McMillan} S. L.~W.,   {Gieles} M.,  2010, \mn@doi
  [\araa] {10.1146/annurev-astro-081309-130834}, \href
  {https://ui.adsabs.harvard.edu/abs/2010ARA&A..48..431P} {48, 431}

\bibitem[\protect\citeauthoryear{{Rastello}, {Amaro-Seoane}, {Arca-Sedda},
  {Capuzzo-Dolcetta}, {Fragione}  \& {Tosta e Melo}}{{Rastello}
  et~al.}{2019}]{Rastello2018}
{Rastello} S.,  {Amaro-Seoane} P.,  {Arca-Sedda} M.,  {Capuzzo-Dolcetta} R.,
  {Fragione} G.,   {Tosta e Melo} I.,  2019, \mn@doi [\mnras]
  {10.1093/mnras/sty3193}, \href
  {https://ui.adsabs.harvard.edu/abs/2019MNRAS.483.1233R} {483, 1233}

\bibitem[\protect\citeauthoryear{{Rastello}, {Mapelli}, {Di Carlo}, {Giacobbo},
  {Santoliquido}, {Spera}, {Ballone}  \& {Iorio}}{{Rastello}
  et~al.}{2020}]{Rastello20}
{Rastello} S.,  {Mapelli} M.,  {Di Carlo} U.~N.,  {Giacobbo} N.,
  {Santoliquido} F.,  {Spera} M.,  {Ballone} A.,   {Iorio} G.,  2020, \mn@doi
  [\mnras] {10.1093/mnras/staa2018}, \href
  {https://ui.adsabs.harvard.edu/abs/2020MNRAS.tmp.2130R} {}

\bibitem[\protect\citeauthoryear{{Rastello}, {Mapelli}, {Di Carlo}, {Iorio},
  {Ballone}, {Giacobbo}, {Santoliquido}  \& {Torniamenti}}{{Rastello}
  et~al.}{2021}]{rastello2021}
{Rastello} S.,  {Mapelli} M.,  {Di Carlo} U.~N.,  {Iorio} G.,  {Ballone} A.,
  {Giacobbo} N.,  {Santoliquido} F.,   {Torniamenti} S.,  2021, \mn@doi
  [\mnras] {10.1093/mnras/stab2355}, \href
  {https://ui.adsabs.harvard.edu/abs/2021MNRAS.507.3612R} {507, 3612}

\bibitem[\protect\citeauthoryear{{Repetto}, {Davies}  \&
  {Sigurdsson}}{{Repetto} et~al.}{2012}]{repetto2012}
{Repetto} S.,  {Davies} M.~B.,   {Sigurdsson} S.,  2012, \mn@doi [\mnras]
  {10.1111/j.1365-2966.2012.21549.x}, \href
  {https://ui.adsabs.harvard.edu/abs/2012MNRAS.425.2799R} {425, 2799}

\bibitem[\protect\citeauthoryear{{Repetto}, {Igoshev}  \& {Nelemans}}{{Repetto}
  et~al.}{2017}]{repetto2017}
{Repetto} S.,  {Igoshev} A.~P.,   {Nelemans} G.,  2017, \mn@doi [\mnras]
  {10.1093/mnras/stx027}, \href
  {https://ui.adsabs.harvard.edu/abs/2017MNRAS.467..298R} {467, 298}

\bibitem[\protect\citeauthoryear{{Rib{\'o}} et~al.,}{{Rib{\'o}}
  et~al.}{2017}]{ribo2017}
{Rib{\'o}} M.,  et~al., 2017, \mn@doi [\apjl] {10.3847/2041-8213/835/2/L33},
  \href {https://ui.adsabs.harvard.edu/abs/2017ApJ...835L..33R} {835, L33}

\bibitem[\protect\citeauthoryear{{Romagnolo}, {Belczynski}, {Klencki},
  {Agrawal}, {Shenar}  \& {Sz{\'e}csi}}{{Romagnolo}
  et~al.}{2022}]{Romagnolo2022}
{Romagnolo} A.,  {Belczynski} K.,  {Klencki} J.,  {Agrawal} P.,  {Shenar} T.,
  {Sz{\'e}csi} D.,  2022, \mn@doi [arXiv e-prints] {10.48550/arXiv.2211.15800},
  \href {https://ui.adsabs.harvard.edu/abs/2022arXiv221115800R} {p.
  arXiv:2211.15800}

\bibitem[\protect\citeauthoryear{{R{\"o}pke} \& {De Marco}}{{R{\"o}pke} \& {De
  Marco}}{2023}]{roepke2022}
{R{\"o}pke} F.~K.,  {De Marco} O.,  2023, \mn@doi [Living Reviews in
  Computational Astrophysics] {10.1007/s41115-023-00017-x}, \href
  {https://ui.adsabs.harvard.edu/abs/2023LRCA....9....2R} {9, 2}

\bibitem[\protect\citeauthoryear{{Sana} et~al.,}{{Sana} et~al.}{2012}]{sana12}
{Sana} H.,  et~al., 2012, \mn@doi [Science] {10.1126/science.1223344}, \href
  {https://ui.adsabs.harvard.edu/abs/2012Sci...337..444S} {337, 444}

\bibitem[\protect\citeauthoryear{{Saracino} et~al.,}{{Saracino}
  et~al.}{2022}]{saracino2022}
{Saracino} S.,  et~al., 2022, \mn@doi [\mnras] {10.1093/mnras/stab3159}, \href
  {https://ui.adsabs.harvard.edu/abs/2022MNRAS.511.2914S} {511, 2914}

\bibitem[\protect\citeauthoryear{{Saracino} et~al.,}{{Saracino}
  et~al.}{2023}]{saracino2023}
{Saracino} S.,  et~al., 2023, \mn@doi [\mnras] {10.1093/mnras/stad764}, \href
  {https://ui.adsabs.harvard.edu/abs/2023MNRAS.521.3162S} {521, 3162}

\bibitem[\protect\citeauthoryear{{Schneider}, {Ohlmann}, {Podsiadlowski},
  {R{\"o}pke}, {Balbus}, {Pakmor}  \& {Springel}}{{Schneider}
  et~al.}{2019}]{schneider2019}
{Schneider} F. R.~N.,  {Ohlmann} S.~T.,  {Podsiadlowski} P.,  {R{\"o}pke}
  F.~K.,  {Balbus} S.~A.,  {Pakmor} R.,   {Springel} V.,  2019, \mn@doi [\nat]
  {10.1038/s41586-019-1621-5}, \href
  {https://ui.adsabs.harvard.edu/abs/2019Natur.574..211S} {574, 211}

\bibitem[\protect\citeauthoryear{{Schneider}, {Ohlmann}, {Podsiadlowski},
  {R{\"o}pke}, {Balbus}  \& {Pakmor}}{{Schneider} et~al.}{2020}]{schneider2020}
{Schneider} F.~R.~N.,  {Ohlmann} S.~T.,  {Podsiadlowski} P.,  {R{\"o}pke}
  F.~K.,  {Balbus} S.~A.,   {Pakmor} R.,  2020, \mn@doi [\mnras]
  {10.1093/mnras/staa1326}, \href
  {https://ui.adsabs.harvard.edu/abs/2020MNRAS.495.2796S} {495, 2796}

\bibitem[\protect\citeauthoryear{{Schoettler}, {Parker}, {Arnold}, {Grimmett},
  {de Bruijne}  \& {Wright}}{{Schoettler} et~al.}{2019}]{Schoettler2019}
{Schoettler} C.,  {Parker} R.~J.,  {Arnold} B.,  {Grimmett} L.~P.,  {de
  Bruijne} J.,   {Wright} N.~J.,  2019, \mn@doi [\mnras]
  {10.1093/mnras/stz1487}, \href
  {https://ui.adsabs.harvard.edu/abs/2019MNRAS.487.4615S} {487, 4615}

\bibitem[\protect\citeauthoryear{{Sgalletta} et~al.,}{{Sgalletta}
  et~al.}{2023}]{sgalletta2023}
{Sgalletta} C.,  et~al., 2023, \mn@doi [arXiv e-prints]
  {10.48550/arXiv.2305.04955}, \href
  {https://ui.adsabs.harvard.edu/abs/2023arXiv230504955S} {p. arXiv:2305.04955}

\bibitem[\protect\citeauthoryear{{Shahaf}, {Mazeh}, {Faigler}  \&
  {Holl}}{{Shahaf} et~al.}{2019}]{shahaf2019}
{Shahaf} S.,  {Mazeh} T.,  {Faigler} S.,   {Holl} B.,  2019, \mn@doi [\mnras]
  {10.1093/mnras/stz1636}, \href
  {https://ui.adsabs.harvard.edu/abs/2019MNRAS.487.5610S} {487, 5610}

\bibitem[\protect\citeauthoryear{{Shahaf}, {Bashi}, {Mazeh}, {Faigler},
  {Arenou}, {El-Badry}  \& {Rix}}{{Shahaf} et~al.}{2023}]{shahaf2023}
{Shahaf} S.,  {Bashi} D.,  {Mazeh} T.,  {Faigler} S.,  {Arenou} F.,  {El-Badry}
  K.,   {Rix} H.~W.,  2023, \mn@doi [\mnras] {10.1093/mnras/stac3290}, \href
  {https://ui.adsabs.harvard.edu/abs/2023MNRAS.518.2991S} {518, 2991}

\bibitem[\protect\citeauthoryear{{Shao} \& {Li}}{{Shao} \&
  {Li}}{2019}]{Shao2019}
{Shao} Y.,  {Li} X.-D.,  2019, \mn@doi [\apj] {10.3847/1538-4357/ab4816}, \href
  {https://ui.adsabs.harvard.edu/abs/2019ApJ...885..151S} {885, 151}

\bibitem[\protect\citeauthoryear{{Shenar} et~al.,}{{Shenar}
  et~al.}{2022}]{shenar2022a}
{Shenar} T.,  et~al., 2022, \mn@doi [Nature Astronomy]
  {10.1038/s41550-022-01730-y}, \href
  {https://ui.adsabs.harvard.edu/abs/2022NatAs...6.1085S} {6, 1085}

\bibitem[\protect\citeauthoryear{{Shikauchi}, {Kumamoto}, {Tanikawa}  \&
  {Fujii}}{{Shikauchi} et~al.}{2020}]{shikauchi2020}
{Shikauchi} M.,  {Kumamoto} J.,  {Tanikawa} A.,   {Fujii} M.~S.,  2020, \mn@doi
  [\pasj] {10.1093/pasj/psaa030}, \href
  {https://ui.adsabs.harvard.edu/abs/2020PASJ...72...45S} {72, 45}

\bibitem[\protect\citeauthoryear{{Shikauchi}, {Tanikawa}  \&
  {Kawanaka}}{{Shikauchi} et~al.}{2022}]{shikauchi2022}
{Shikauchi} M.,  {Tanikawa} A.,   {Kawanaka} N.,  2022, \mn@doi [\apj]
  {10.3847/1538-4357/ac5329}, \href
  {https://ui.adsabs.harvard.edu/abs/2022ApJ...928...13S} {928, 13}

\bibitem[\protect\citeauthoryear{{Shikauchi}, {Tsuna}, {Tanikawa}  \&
  {Kawanaka}}{{Shikauchi} et~al.}{2023}]{shikauchi23}
{Shikauchi} M.,  {Tsuna} D.,  {Tanikawa} A.,   {Kawanaka} N.,  2023, \mn@doi
  [arXiv e-prints] {10.48550/arXiv.2301.07207}, \href
  {https://ui.adsabs.harvard.edu/abs/2023arXiv230107207S} {p. arXiv:2301.07207}

\bibitem[\protect\citeauthoryear{{Stiefel}, {Rheinboldt}, {Rheinboldt}  \&
  {Hagger}}{{Stiefel} et~al.}{1965}]{stiefel65}
{Stiefel} E.~L.,  {Rheinboldt} W.~C.,  {Rheinboldt} C.~J.,   {Hagger} H.~J.,
  1965, \mn@doi [Physics Today] {10.1063/1.3047091}, \href
  {https://ui.adsabs.harvard.edu/abs/1965PhT....18a.110S} {18, 110}

\bibitem[\protect\citeauthoryear{{Tanikawa}, {Cary}, {Shikauchi}, {Wang}  \&
  {Fujii}}{{Tanikawa} et~al.}{2023a}]{tanikawa2023}
{Tanikawa} A.,  {Cary} S.,  {Shikauchi} M.,  {Wang} L.,   {Fujii} M.~S.,
  2023a, \mn@doi [arXiv e-prints] {10.48550/arXiv.2303.05743}, \href
  {https://ui.adsabs.harvard.edu/abs/2023arXiv230305743T} {p. arXiv:2303.05743}

\bibitem[\protect\citeauthoryear{{Tanikawa}, {Hattori}, {Kawanaka}, {Kinugawa},
  {Shikauchi}  \& {Tsuna}}{{Tanikawa} et~al.}{2023b}]{tanikawa2023bh}
{Tanikawa} A.,  {Hattori} K.,  {Kawanaka} N.,  {Kinugawa} T.,  {Shikauchi} M.,
   {Tsuna} D.,  2023b, \mn@doi [\apj] {10.3847/1538-4357/acbf36}, \href
  {https://ui.adsabs.harvard.edu/abs/2023ApJ...946...79T} {946, 79}

\bibitem[\protect\citeauthoryear{{Thompson} et~al.,}{{Thompson}
  et~al.}{2019}]{thompson2019}
{Thompson} T.~A.,  et~al., 2019, \mn@doi [Science] {10.1126/science.aau4005},
  \href {https://ui.adsabs.harvard.edu/abs/2019Sci...366..637T} {366, 637}

\bibitem[\protect\citeauthoryear{{Tokovinin} \& {Moe}}{{Tokovinin} \&
  {Moe}}{2020}]{tokovinin2020}
{Tokovinin} A.,  {Moe} M.,  2020, \mn@doi [\mnras] {10.1093/mnras/stz3299},
  \href {https://ui.adsabs.harvard.edu/abs/2020MNRAS.491.5158T} {491, 5158}

\bibitem[\protect\citeauthoryear{{Torniamenti}, {Ballone}, {Mapelli},
  {Gaspari}, {Di Carlo}, {Rastello}, {Giacobbo}  \& {Pasquato}}{{Torniamenti}
  et~al.}{2021}]{torniamenti2021}
{Torniamenti} S.,  {Ballone} A.,  {Mapelli} M.,  {Gaspari} N.,  {Di Carlo}
  U.~N.,  {Rastello} S.,  {Giacobbo} N.,   {Pasquato} M.,  2021, \mn@doi
  [\mnras] {10.1093/mnras/stab2238}, \href
  {https://ui.adsabs.harvard.edu/abs/2021MNRAS.507.2253T} {507, 2253}

\bibitem[\protect\citeauthoryear{{Torniamenti}, {Rastello}, {Mapelli}, {Di
  Carlo}, {Ballone}  \& {Pasquato}}{{Torniamenti}
  et~al.}{2022}]{torniamenti2022}
{Torniamenti} S.,  {Rastello} S.,  {Mapelli} M.,  {Di Carlo} U.~N.,  {Ballone}
  A.,   {Pasquato} M.,  2022, \mn@doi [\mnras] {10.1093/mnras/stac2841}, \href
  {https://ui.adsabs.harvard.edu/abs/2022MNRAS.517.2953T} {517, 2953}

\bibitem[\protect\citeauthoryear{{Torniamenti}, {Gieles}, {Penoyre},
  {Jerabkova}, {Wang}  \& {Anders}}{{Torniamenti}
  et~al.}{2023}]{torniamenti2023}
{Torniamenti} S.,  {Gieles} M.,  {Penoyre} Z.,  {Jerabkova} T.,  {Wang} L.,
  {Anders} F.,  2023, \mn@doi [arXiv e-prints] {10.48550/arXiv.2303.10188},
  \href {https://ui.adsabs.harvard.edu/abs/2023arXiv230310188T} {p.
  arXiv:2303.10188}

\bibitem[\protect\citeauthoryear{{Trani}, {Rastello}, {Di Carlo},
  {Santoliquido}, {Tanikawa}  \& {Mapelli}}{{Trani} et~al.}{2022}]{Trani2022}
{Trani} A.~A.,  {Rastello} S.,  {Di Carlo} U.~N.,  {Santoliquido} F.,
  {Tanikawa} A.,   {Mapelli} M.,  2022, \mn@doi [\mnras]
  {10.1093/mnras/stac122}, \href
  {https://ui.adsabs.harvard.edu/abs/2022MNRAS.511.1362T} {511, 1362}

\bibitem[\protect\citeauthoryear{{Trimble} \& {Thorne}}{{Trimble} \&
  {Thorne}}{1969}]{trimble1969}
{Trimble} V.~L.,  {Thorne} K.~S.,  1969, \mn@doi [\apj] {10.1086/150032}, \href
  {https://ui.adsabs.harvard.edu/abs/1969ApJ...156.1013T} {156, 1013}

\bibitem[\protect\citeauthoryear{{Wang}, {Spurzem}, {Aarseth}, {Nitadori},
  {Berczik}, {Kouwenhoven}  \& {Naab}}{{Wang} et~al.}{2015}]{wang15}
{Wang} L.,  {Spurzem} R.,  {Aarseth} S.,  {Nitadori} K.,  {Berczik} P.,
  {Kouwenhoven} M.~B.~N.,   {Naab} T.,  2015, \mn@doi [\mnras]
  {10.1093/mnras/stv817}, \href
  {https://ui.adsabs.harvard.edu/abs/2015MNRAS.450.4070W} {450, 4070}

\bibitem[\protect\citeauthoryear{{Webbink}}{{Webbink}}{1984}]{webbink84}
{Webbink} R.~F.,  1984, \mn@doi [\apj] {10.1086/161701}, \href
  {https://ui.adsabs.harvard.edu/abs/1984ApJ...277..355W} {277, 355}

\bibitem[\protect\citeauthoryear{{Wiktorowicz}, {Lu}, {Wyrzykowski}, {Zhang},
  {Liu}, {Justham}  \& {Belczynski}}{{Wiktorowicz}
  et~al.}{2020}]{Wiktorowicz2020}
{Wiktorowicz} G.,  {Lu} Y.,  {Wyrzykowski} {\L}.,  {Zhang} H.,  {Liu} J.,
  {Justham} S.,   {Belczynski} K.,  2020, \mn@doi [\apj]
  {10.3847/1538-4357/abc699}, \href
  {https://ui.adsabs.harvard.edu/abs/2020ApJ...905..134W} {905, 134}

\bibitem[\protect\citeauthoryear{{Yalinewich}, {Beniamini}, {Hotokezaka}  \&
  {Zhu}}{{Yalinewich} et~al.}{2018}]{yalinewich2018}
{Yalinewich} A.,  {Beniamini} P.,  {Hotokezaka} K.,   {Zhu} W.,  2018, \mn@doi
  [\mnras] {10.1093/mnras/sty2327}, \href
  {https://ui.adsabs.harvard.edu/abs/2018MNRAS.481..930Y} {481, 930}

\bibitem[\protect\citeauthoryear{{Yamaguchi}, {Kawanaka}, {Bulik}  \&
  {Piran}}{{Yamaguchi} et~al.}{2018}]{yamaguchi2018}
{Yamaguchi} M.~S.,  {Kawanaka} N.,  {Bulik} T.,   {Piran} T.,  2018, \mn@doi
  [\apj] {10.3847/1538-4357/aac5ec}, \href
  {https://ui.adsabs.harvard.edu/abs/2018ApJ...861...21Y} {861, 21}

\makeatother
\end{thebibliography}




\appendix 

\section{Population of Retained BH-MS binaries}\label{sec:app}

\begin{figure}
    \centering      
\includegraphics[width=0.45\textwidth]{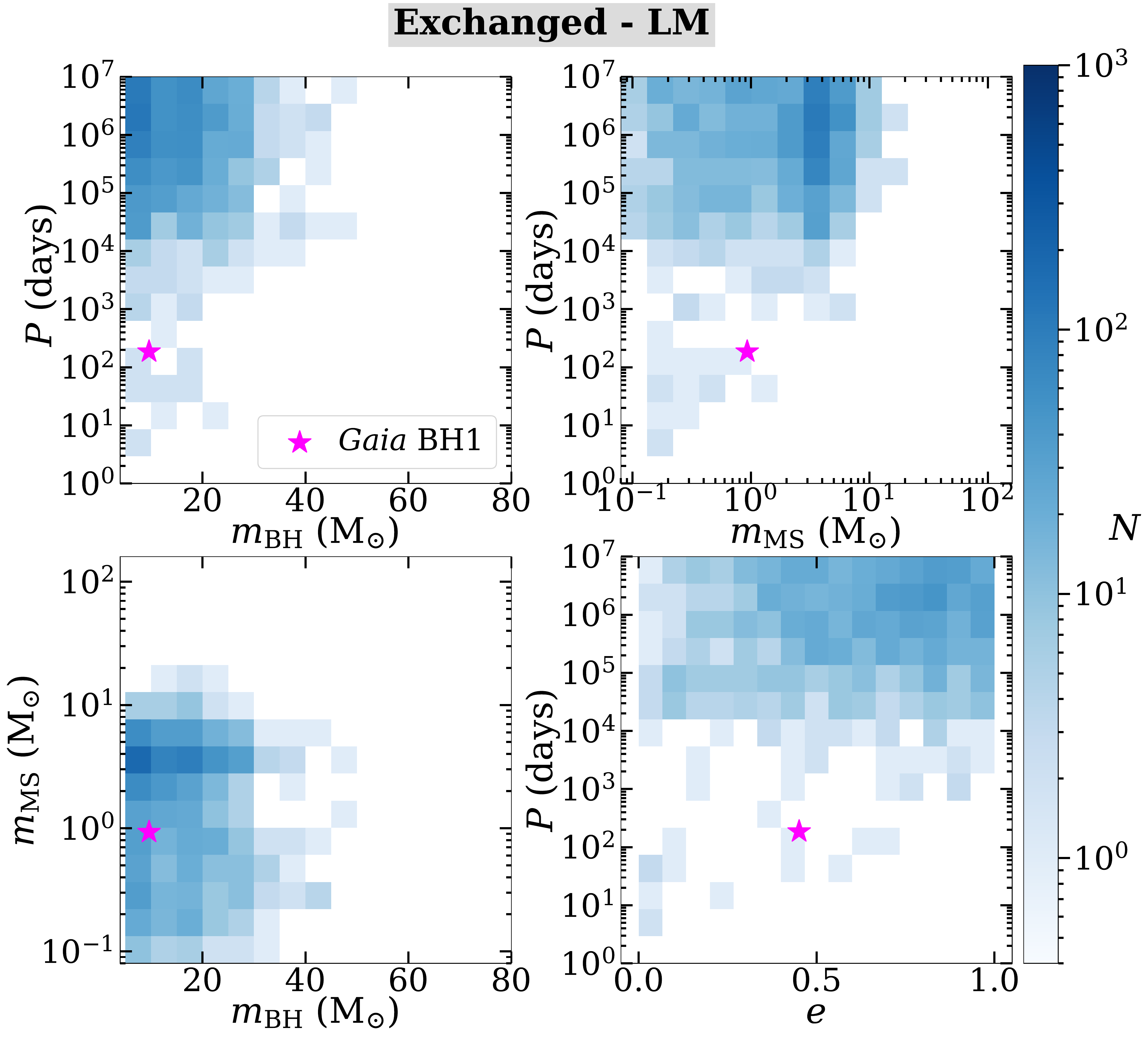}
    \caption{Same as Fig.~\ref{fig:low-ej-ex} but for 
    retained BH-MS in LM clusters.}
    \label{fig:ret-lm}
\end{figure}

\begin{figure}
    \centering
    \includegraphics[width=0.45\textwidth]{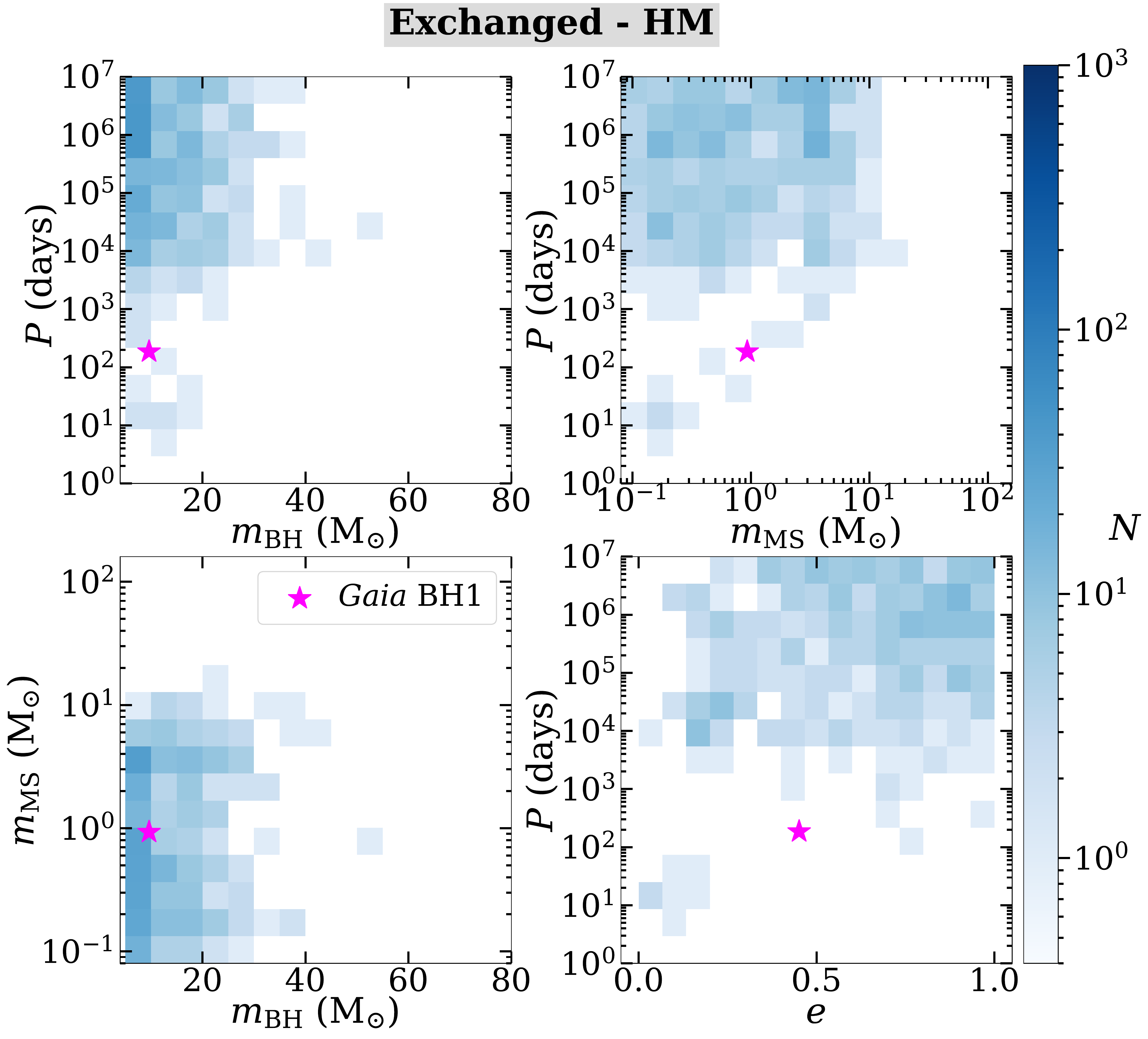}
    \caption{Same as Fig.~\ref{fig:ret-lm} but for 
    HM clusters.}
    \label{fig:ret_hm} 
\end{figure}

\begin{figure}    
\centering  
    \includegraphics[width=0.45\textwidth]{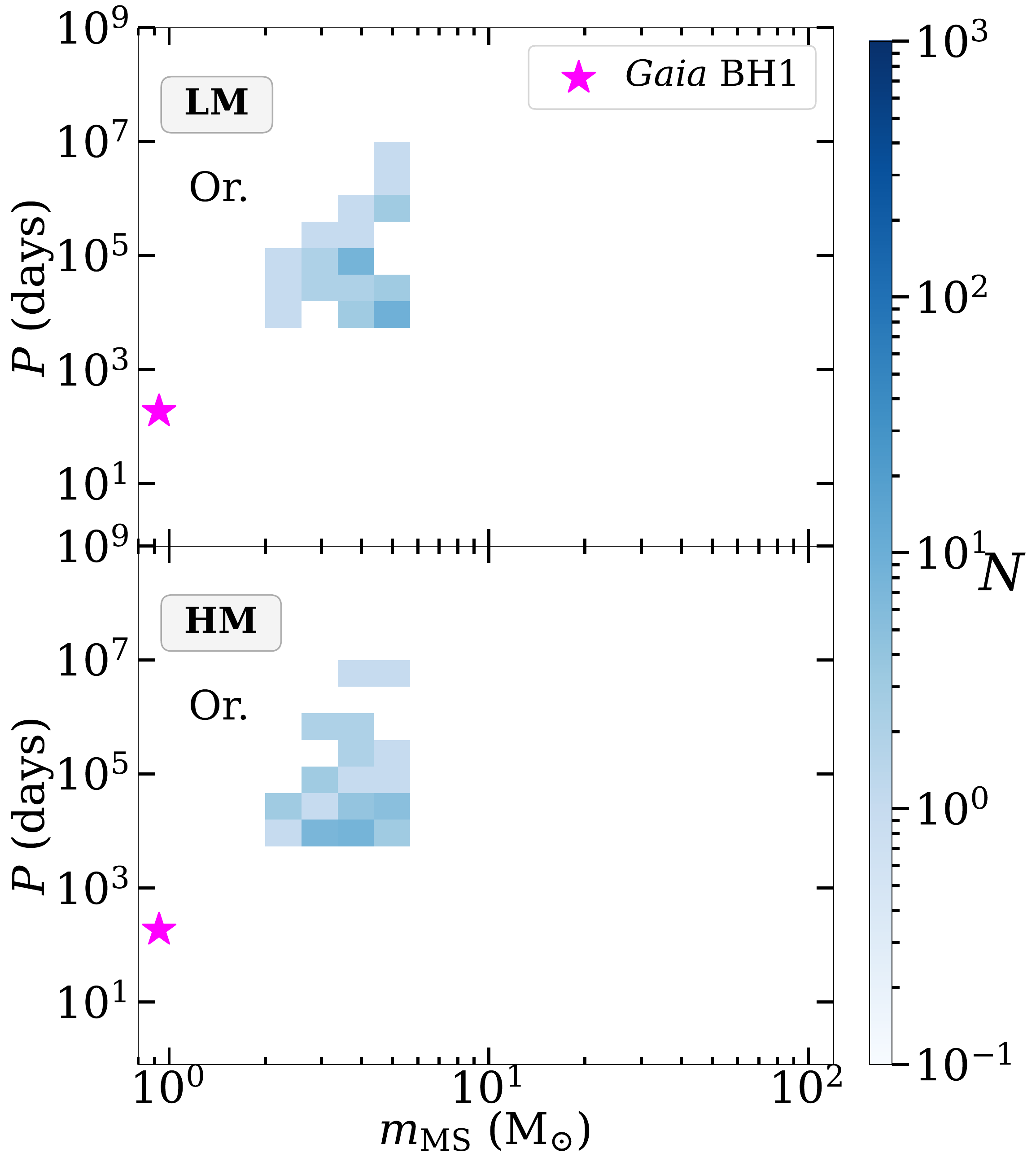} 
    \caption{Same as Fig.~\ref{fig:orig_iso} but for 
    retained original BH-MS binaries.}
    \label{fig:ret_orig}
\end{figure}

At the end of the simulations ($\approx \ 100 $ Myr), only $\approx 7 \%$ of the formed BH-MS are still within the HM and LM clusters. All the others have been either ejected or destroyed or simply left the MS phase.
 In both LM and HM clusters most of such retained BH-MS binaries are exchanged (97\% and 90\% respectively), whilst the remaining binaries are original. Since our simulations last for only 
100 Myr, it is likely that even the retained  BH-MS binary systems are ejected or evaporate from their parent cluster in  
later evolutionary stages. Hence, here we discuss the properties of BH-MS stars that remain bound to their parent cluster at the end of the simulations (Figs.~\ref{fig:ret-lm}, \ref{fig:ret_hm}, and \ref{fig:ret_orig}).
As expected the population of retained BH-MS at 100 Myr (both exchanged and original) in both LM and HM clusters host mostly 
low-mass MS stars and long-period ($P \geq 1000$ days) binaries that have not been destroyed or ejected yet by dynamical encounters within the parent cluster (see also \citealp{torniamenti2023}).
LM clusters retain one \bh{}-like binary (according to the selection window adopted in Sec.~\ref{sec:eff}). In contrast, none of the BH-MS binaries retained in HM clusters are \bh{}-like systems.


\bsp	
\label{lastpage}
\end{document}